\documentclass[epj]{svjour}
\usepackage{amssymb}
\usepackage{graphics}

\begin{document}

\title{Phonon driven transport in amorphous semiconductors}
\subtitle{: Transition probabilities}
\author{Ming-Liang Zhang and D. A. Drabold}
\institute{Department of Physics and Astronomy, Ohio University, Athens, Ohio 45701 }
\date{Received: date / Revised version: date}

\abstract{
Starting from Holstein's work on small polaron hopping, the evolution
equations for localized and extended states in the presence of atomic
vibrations are systematically derived for an amorphous semiconductor. The
transition probabilities are obtained for transitions between all combinations
of localized and extended states. For any transition process involving a
localized state, the activation energy is not simply the energy difference
between the final and initial states; the reorganization energy of atomic
configuration is also included as an important part of the activation
energy (Marcus form). The activation energy for the transitions between
localized states decreases with rising temperature and leads to the
Meyer-Neldel rule. The predicted Meyer-Neldel temperatures are consistent with
observations in several materials. The computed field-dependence of
conductivity agrees with experimental data. The present work suggests that the
upper temperature limit of variable range hopping is proportional to the
frequency of first peak of phonon spectrum. We have also improved the
description of the photocurrent decay at low temperatures. Analysis of the
transition probability from an extended state to a localized state suggests
that there exists a short-lifetime belt of extended states inside the
conduction band or valence band.
}
\maketitle

\section{Introduction}

\label{intro} In the last 50 years, transport properties in amorphous
semiconductors have been intensely studied\cite{eve,dyr,bern,ram,pl,les}.
Miller-Abrahams (MA) theory\cite{mil} and variable range hopping (VRH) \cite%
{md} are frequently used to fit dc conductivity data. For localized tail
states close to a mobility edge, `phonon induced delocalization'\cite{kik}
plays an important role in transport. In addition, exciton hopping among
localized states is suggested as the mechanism of photoluminescence in a
quantum well\cite{gou}. The Meyer-Neldel rule has been deduced from the
shift of Fermi level\cite{Over}, from multi-excitation entropy\cite%
{Yelon90,Yelon06} and from other perspectives\cite{wan,cra,Emin08}.

In the MA theory of dc conductivity\cite{mil}, the polarization of the
network by impurity atoms and by carriers in localized tail states was
neglected. The transitions between localized states were induced by
single-phonon absorption or emission. Subsequent research on transient
photocurrent decay\cite{sch,ore,kris,mon} adopted a parameterized MA
transition probability and therefore inherited the single-phonon features of
MA theory. However in other electronic hopping processes, the polarization
of the environment by moving electrons plays an important role. Electron
transfer in polar solvents, electron transfer inside large molecules\cite{marcu} and
polaron diffusion in a molecular crystal\cite{Hol1,Hol2} are relevant examples.
The transition probability $W_{LL}$ between two sites in a thermally
activated process is given by the Marcus formula\cite{marcu}:%
\begin{equation}
W_{LL}=\nu_{LL}e^{-E_{a}^{LL}/k_{B}T}, ~~E_{a}^{LL}=\frac{\lambda_{LL}}{4}(1+%
\frac{\Delta G_{LL}^{0}}{\lambda_{LL}})^{2},  \label{ea}
\end{equation}
where $\nu_{LL}$ has the dimension of frequency that characterizes a
specific hopping process. Here, $E_{a}^{LL}$ is the temperature-dependent
activation energy, $\lambda_{LL}$ is the reorganization energy, $\Delta
G_{LL}^{0}$ is the energy difference between the final state and the initial
state\cite{marcu}. (\ref{ea}) has been established for both electron transfer%
\cite{marcu} and small polaron hopping\cite{rei}. The mathematical form of
Holstein's work for 1d molecular crystals is quite flexible and can be used
for 3d materials with slight modifications\cite%
{Emin74,Emin75,Emin76,Emin77,Emin91}. Emin applied small polaron theory to transport properties in amorphous semiconductors, and assumed that the
static displacements of atoms induced by electron-phonon (e-ph) interaction
caused carrier self-trapping on a single atom\cite{com}. The effect of
static disorder was taken into account by replacing a fixed transfer
integral with a distribution of. He found that the static disorder reduces the
strength of the electron-lattice coupling needed to stabilize global
small-polaron fomation\cite{Emin94}. B\"{o}ttger and Bryksin summarized\cite%
{bot} a wide range literature on hopping conduction in solids before 1984.

By estimating the size of various contributions, we show in Sec.II that (1)
static disorder is more important than the static displacement induced by
e-ph interaction, so that the carriers are localized at the band tails by
the static disorder\cite{md,and}; (2) the static displacement induced by the
e-ph interaction in a localized state is much larger than the vibrational
amplitude of the atoms, so that any transition involving localized state(s)
must be a multi-phonon process. In a semiconductor, the mid-gap states and
localized band tail states are the low-lying excited states (most important
for transport) at moderate temperature\cite{md,and}. To simplify the
problem, let us leave aside the mid-gap states (induced by impurity atoms,
dangling bonds and other defects) and restrict attention to localized tail
states (induced by topological disorder) and extended states only. Mid-gap
states are well treated by other methods\cite{md}.

The first aim of this work is to extend Holstein's work\cite{Hol1,Hol2} to
amorphous semiconductors, and to derive the equations of time evolution for
localized tail states and extended states in the presence of atomic
vibrations.
If there are only two localized states in system, the evolution equations
reduce to the Marcus theory of electron transfer. If there is one localized
state and one extended state, the evolution equations are simplified to
Kramers' problem of particle escape-capture. The short-time
solution of the evolution equations can be used to
compute a spatially averaged current density\cite{kubo} \textit{i.e.} conductivity
(we will report this in a forthcoming paper).

The second aim of this paper is to estimate the transition probabilities of
four elementary processes: (i) transition from a localized tail state to
another localized tail state (LL); (ii) transition from a localized tail
state to an extended state (LE); (iii) transition from an extended state to
a localized tail state (EL) and (iv) transition from an extended state to
another extended state (EE). The distribution functions of carriers in
localized states and in extended states satisfy two coupled generalized
Boltzmann equations. The transition probabilities of LL, LE, EL and EE
transitions obtained are necessary input for these two equations. Then
transport properties could be computed from these Boltzmann equations. In
this paper, we will not pursue this approach, and instead only estimate
conductivity from the intuitive picture of hopping and mean free path. The
present work on four transitions illuminates the physical processes in
transport which are obscured in {\textit{ab initio}} estimations of the
conductivity\cite{kubo}.

\begin{figure}[tbp]
\resizebox{0.5\textwidth}{!}{  \includegraphics{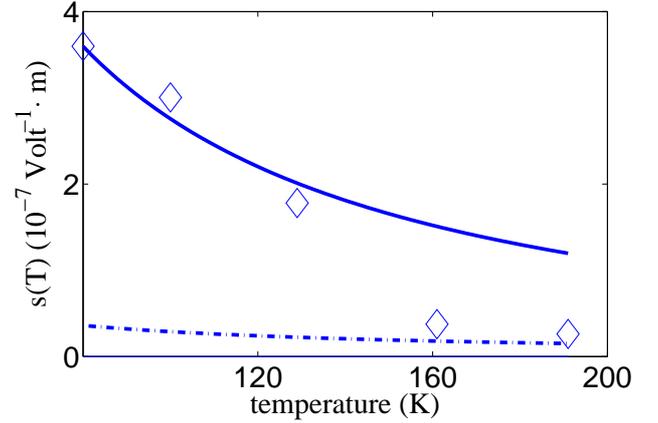}
} 
\caption{The slope $s(T)$ (see text) for a-Ge: diamonds are data\protect\cite%
{ell}. The dashed line is expected from the simple
picture that the potential energy drops along the inverse direction
of field for electrons\protect\cite{bot,fren}. The solid line is calculated from (\protect\ref{fcon}). To estimate
the average over various localized states, we take $\Delta G_{LL}=k_{B}T$.}
\label{aGe}
\end{figure}

The third aim is to (i) check if the new results describe experiments better
than previous theories; (ii) establish new relations for existing data and
(iii) predict new observable phenomenon. In LL, LE and EL transitions, the
frequency of first peak $\bar{\nu}$ of the phonon spectrum supplies energy
separating different transport behavior. In the high temperature regime ($T>h%
\bar{\nu}/k_{B}$), static displacements induced by the e-ph interaction
require configurational reorganization in a transition involving localized
state(s). A Marcus type transition rate is found at `very' high temperature (%
$T>2.5h\bar{\nu}/k_{B}$). The decreasing reorganization energy with rising
temperature leads to the Meyer-Neldel rule: this is a special dynamic
realization of the multi-excitation entropy model\cite{Yelon06}. The
predicted Meyer-Neldel temperature is satisfied in various materials (cf.
Table \ref{tab:tmn}). At low temperature ($T<h\bar{\nu}/10k_{B}$), the
atomic static displacements induced by e-ph interaction reduce the transfer
integral, and a few phonons may be involved in LL, LE and EL transitions. For LL
transition: VRH may be more effective than the transition between
neighboring localized states. The upper temperature limit of VRH is found to be
proportional to $\bar{\nu}$. This new relation is compared to existing
experimental data in Figure \ref{peak}. In the intermediate range($h\bar{\nu}%
/10k_{B}<T<h\bar{\nu}/k_{B}$), the well-known non-Arrhenius and non-VRH
behavior appears naturally in the present framework. Comparing with previous
models, the present work improves the description of the field-dependent
conductivity, cf. Figure \ref{aGe} and Figure \ref{Frenkel}. The predicted
time decay of the photocurrent is compared with observations in a-Si:H and
a-As$_{2}$Se$_{3}$ and is improved, compared to previous models at low
temperature, cf. Figure \ref{transient}. For EL transition, we suggest that
there exists a short-lifetime belt of extended states inside conduction band
or valence band (cf. Figure \ref{belt}).

\section{Evolution of states driven by vibrations}

\label{ev1}

\subsection{Single-electron approximation and evolution equations}

\label{sp} 
For definiteness, we consider electrons in the conduction band of an
amorphous semiconductor. For carriers in mid-gap states and holes in the
valence band, we need only modify the notation slightly. For the case of
intrinsic and lightly doped n-type semiconductors, the number of electrons
is much smaller than the number of localized states. The correlation between
electrons in a hopping process and the screening caused by these electrons
may be neglected. Essentially we have a single particle problem: one
electron interacts with localized tail states and extended states in the
conduction band.

Consider then, one electron in an amorphous solid with $\mathcal{N}$ atoms.
For temperature $T$ well below the melting point $T_{m}$, the atoms execute
small harmonic oscillations around their equilibrium positions $\{\mathcal{R}%
_{\mathbf{n}}\}$: $\mathbf{W}_{\mathbf{n}}=\mathcal{R}_{\mathbf{n}}+\mathbf{u%
}^{\mathbf{n}}$, where $\{\mathbf{W}_{\mathbf{n}}\}$ and $\{\mathbf{u}^{%
\mathbf{n}}\}$ are the instantaneous positions and vibrational
displacements. The wave function $\psi(\mathbf{r},\{\mathbf{u}^{\mathbf{n}%
}\};t)$ is then a function of the vibrational displacements of all atoms and
of the coordinates $\mathbf{r}$ of the electron. The Hamiltonian $H_{1}$ of
the \textquotedblleft one electron + many nuclei\textquotedblright\ system
may be separated into: $H_{1}=h_{e}+h_{v}$, where:
\begin{equation}
h_{e}=\frac{-\hbar^{2}}{2m}\nabla^{2}+\sum_{\mathbf{n}=1}^{\mathcal{N}}U(%
\mathbf{r},\mathcal{R}_{\mathbf{n}},\mathbf{u}^{\mathbf{n}}),  \label{hes}
\end{equation}
is the single-electron Hamiltonian including the vibrations. In $h_{e}$ put $%
\{\mathbf{u}^{\mathbf{n}}=0\}$, and one obtains the hamiltonian $h_{a}$ for
an electron moving in a network with static disorder. Let%
\begin{equation}
h_{v}=\sum_{j}-\frac{\hbar^{2}}{2M_{j}}\nabla_{j}^{2}+\frac{1}{2}\sum
_{jk}k_{jk}x_{j}x_{k},  \label{hvb}
\end{equation}
be the vibrational Hamiltonian, where $(k_{jk})$ is the matrix of force
constants. To simplify, we rename $\{\mathbf{u}^{\mathbf{n}},\mathbf{n}%
=1,2,\cdots,\mathcal{N}\}$ as $\{x_{j},$ $j=1,2,\cdots,3\mathcal{N}\}$. The
evolution of $\psi(\mathbf{r},\{\mathbf{u}^{\mathbf{n}}\};t)$ is given by:%
\begin{equation}
i\hbar\frac{\partial\psi(\mathbf{r},\{\mathbf{u}^{\mathbf{n}}\};t)}{\partial
t}=H_{1}\psi(\mathbf{r},\{\mathbf{u}^{\mathbf{n}}\};t).  \label{fs1}
\end{equation}

The Hilbert space of $h_{e}$ is spanned by the union of localized states $%
\{\phi_{A_{1}}\}$ and the extended states $\{\xi_{B_{1}}\}$. \newline
$\psi (\mathbf{r},\{\mathbf{u}^{\mathbf{n}}\};t)$ may be expanded as%
\begin{equation}
\psi(\mathbf{r},x_{1},\cdots,x_{3\mathcal{N}};t)=\sum_{A_{1}}a_{A_{1}}%
\phi_{A_{1}}+\sum_{B_{1}}b_{B_{1}}\xi_{B_{1}},  \label{fw1}
\end{equation}
where $a_{A_{1}}$ is the probability amplitude at moment $t$ that the
electron is in localized state $A_{1}$ while the displacements of the nuclei
are $\{x_{j},$ $j=1,2,\cdots3\mathcal{N}\}$; $b_{B_{1}}$ is the amplitude at
moment $t$ that the electron is in extended state $B_{1}$ while the
displacements of the nuclei are $\{x_{j},$ $j=1,2,\cdots3\mathcal{N}\}$. To
get the evolution equations for $a_{A_{1}}$ and $b_{B_{1}}$, one substitutes
(\ref{fw1}) into (\ref{fs1}), and separately applies $\int
d^{3}r\phi_{A_{2}}^{\ast}$ and $\int d^{3}r\xi_{B_{2}}^{\ast}$ to both sides
of the equation\cite{Hol1,Hol2}. After some approximations (cf. Appendix A),
the evolution equations are simplified to:
\begin{equation}
\{i\hbar\frac{\partial}{\partial t}-E_{A_{2}}-h_{v}\}a_{A_{2}}=%
\sum_{A_{1}}J_{A_{2}A_{1}}a_{A_{1}}+\sum_{B_{1}}K_{A_{2}B_{1}}^{%
\prime}b_{B_{1}},  \label{s11}
\end{equation}
and
\begin{equation}
\{i\hbar\frac{\partial}{\partial t}-E_{B_{2}}-h_{v}\}b_{B_{2}}=%
\sum_{A_{1}}J_{B_{2}A_{1}}^{\prime}a_{A_{1}}+%
\sum_{B_{1}}K_{B_{2}B_{1}}b_{B_{1}},  \label{s22}
\end{equation}
where $E_{B_{2}}$ is the energy eigenvalue of $h_{a}$ corresponding to
extended state $\xi_{B_{2}}$, and
\begin{equation}
E_{A_{1}}(\{x_{p_{A_{1}}}^{A_{1}}\})=E_{A_{1}}^{0}-\sum_{p_{A_{1}}\mathbf{%
\in }D_{A_{1}}}d_{p_{A_{1}}}x_{p_{A_{1}}}^{A_{1}}  \label{ea1}
\end{equation}
is the energy of localized state $\phi_{A_{1}}$ to the first order of e-ph
interaction, where $d_{p_{A_{1}}}=\int d\mathbf{r}|\phi_{A_{1}}^{0}(\mathbf{r%
},\{\mathcal{R}_{\mathbf{n}}\})|^{2}\partial U/\partial X_{p_{A_{1}}}$. $%
E_{A_{1}}^{0}$ and $\phi_{A_{1}}^{0}$ are the corresponding eigenvalue and
eigenfunction of $h_{a}$ and $\phi_{A_{1}}$ is correction of $%
\phi_{A_{1}}^{0}$ to the first order of e-ph interaction.

The transfer integral
\begin{equation}
J_{A_{2}A_{1}}=\int d^{3}r\phi _{A_{2}}^{\ast }\sum_{\mathbf{p}\in
D_{A_{2}}}U(r-\mathcal{R}_{\mathbf{p}},\mathbf{u}^{\mathbf{p}})\phi _{A_{1}},
\label{tra1}
\end{equation}%
induces transitions from $\phi _{A_{1}}$ to $\phi _{A_{2}}$.
Similarly $J_{A_{1}A_{2}}$ induces transitions from $A_{2}$ to $A_{1}$, and
is due to the attraction on the electron by the atoms in $D_{A_{1}}$. From
definition (\ref{tra1}), no simple relation exists between $J_{A_{2}A_{1}}$
and $J_{A_{1}A_{2}}$. This is because: (i) the number of atoms in $D_{A_{1}}$
may be different to that in $D_{A_{2}}$; (ii) even the numbers of atoms are
the same, the atomic configurations can be different due to the topological
disorder in amorphous materials. This is in contrast with the situation of
small polarons in a crystal\cite{Hol2}: $J_{A_{2}A_{1}}=(J_{A_{1}A_{2}})^{%
\ast }$, where two lattice sites are identical. The non-Hermiticity of $%
J_{A_{2}A_{1}}$ comes from the asymmetric potential energy partition (\ref{sl}%
) for well localized states. The transition probability from a localized
state to another is different with its inverse process.

The e-ph interaction
\begin{equation}
K_{A_{2}B_{1}}^{\prime }=\sum_{j}x_{j}\int d^{3}r\phi _{A_{2}}^{\ast }\frac{%
\partial U}{\partial X_{j}}\xi _{B_{1}},  \label{ept}
\end{equation}%
is a linear function of atomic displacements $x_{j}$. It causes EL
transitions from extended states to localized states. LE transition from a
localized tail state in region $D_{A_{1}}$ to an extended state (LE) is
induced by the transfer integral:%
\begin{equation}
J_{B_{2}A_{1}}^{\prime }=\sum_{\mathbf{p}\notin D_{A_{1}}}\int d^{3}r\xi
_{B_{2}}^{\ast }U(r-\mathcal{R}_{\mathbf{p}},\mathbf{u}^{\mathbf{p}})\phi
_{A_{1}}~,  \label{tra2}
\end{equation}%
not by the e-ph interaction $K_{A_{2}B_{1}}^{\prime }$.
Later we neglect the dependence of $J_{B_{2}A_{1}}^{\prime }$ on the
displacements of atoms and only view $J_{B_{2}A_{1}}^{\prime }$ as function
of $\xi _{A_{1}}$ only. The EE transition between two extended states $\xi
_{B_{1}}$ and $\xi _{B_{2}}$ is caused by e-ph interaction:%
\[
K_{B_{2}B_{1}}=\sum_{j}x_{j}\int d^{3}r\xi _{B_{2}}^{\ast }\frac{\partial U}{%
\partial X_{j}}\xi _{B_{1}},
\]%
\begin{equation}
~~~K_{B_{2}B_{1}}=(K_{B_{1}B_{2}})^{\ast }~.  \label{ep0}
\end{equation}%
It is almost identical to scattering between two Bloch states by the e-ph
interaction. In contrast with LL, LE and EL transitions, transition $\xi
_{B_{1}}\rightarrow \xi _{B_{2}}$ and its inverse process $\xi
_{B_{2}}\rightarrow \xi _{B_{1}}$ are coupled by the \textit{same}
interaction, as illustrated in (\ref{ep0}). The transition probabilities of
the two processes are equal. The numerical magnitude of the coupling
parameters of the four transitions are estimated in Appendix B.

\subsection{Reformulation using Normal Coordinates}

\label{nc}

As usual, it is convenient to convert $\{x_{k}\}$ to normal coordinates\cite%
{dra,tafn} $\{\Theta \}$,%
\begin{equation}
x_{k}=\sum_{\alpha }\Delta _{k\alpha }\Theta _{\alpha },~~~(\Delta
^{T}k\Delta )_{\beta \alpha }=\delta _{\alpha \beta }M_{\alpha }\omega
_{\alpha }^{2},  \label{nor1}
\end{equation}%
\[
\alpha =1,2,\cdots ,3\mathcal{N},
\]%
where $\Delta _{k\alpha }$ is the minor of the determinant $\left\vert
k_{ik}-\omega ^{2}M_{i}\delta _{ik}\right\vert =0$, $\Delta ^{T}$ is the
transpose matrix of the matrix $(\Delta _{k\alpha })$. The two coupling
constants in (\ref{ept}) and (\ref{ep0}) which involve e-ph interaction are
expressed as:%
\begin{equation}
K_{A_{2}B_{1}}^{\prime }=\sum_{\alpha }\Theta _{\alpha
}K_{A_{2}B_{1}}^{\prime \alpha },  \label{kp}
\end{equation}%
\[
~~K_{A_{2}B_{1}}^{\prime \alpha }=\sum_{j}\Delta _{j\alpha }\int d^{3}r\phi
_{A_{2}}^{\ast }\frac{\partial U}{\partial X_{j}}\xi _{B_{1}},
\]%
and
\[
K_{B_{2}B_{1}}=\sum_{\alpha }\Theta _{\alpha }K_{B_{2}B_{1}}^{\alpha },
\]%
\begin{equation}
~~K_{B_{2}B_{1}}^{\alpha }=\sum_{j}\Delta _{j\alpha }\int d^{3}r\xi
_{B_{2}}^{\ast }\frac{\partial U}{\partial X_{j}}\xi _{B_{1}},  \label{k}
\end{equation}%
where $K_{A_{2}B_{1}}^{\prime \alpha }$ and $K_{A_{2}B_{1}}^{\alpha }$have
the dimension of force. (\ref{s11}) and (\ref{s22}) become:%
\[
(i\hbar \frac{\partial }{\partial t}-h_{A_{2}})a_{A_{2}}(\cdots \Theta
_{\alpha }\cdots ;t)
\]%
\begin{equation}
=\sum_{A_{1}}J_{A_{2}A_{1}}a_{A_{1}}+\sum_{B_{1}}K_{A_{2}B_{1}}^{\prime
}b_{B_{1}},  \label{1s}
\end{equation}%
and
\[
(i\hbar \frac{\partial }{\partial t}-h_{B_{2}})b_{B_{2}}(\cdots \Theta
_{\alpha }\cdots ;t)
\]%
\begin{equation}
=\sum_{A_{1}}J_{B_{2}A_{1}}^{\prime
}a_{A_{1}}+\sum_{B_{1}}K_{B_{2}B_{1}}b_{B_{1}},  \label{2s}
\end{equation}%
where $h_{A_{1}}=E_{A_{1}}+h_{v}$ describes the polarization of the
amorphous network caused by an electron in localized state $\phi _{A_{1}}$
mediated by the e-ph coupling, and $h_{B_{2}}=E_{B_{2}}+h_{v}$. $%
a_{A_{2}}(\cdots \Theta _{\alpha }\cdots ;t)$ is the probability amplitude
at moment $t$ that the electron is in localized state $\phi _{A_{2}}$ while
the vibrational state of the atoms is given by normal coordinates $\{\Theta
_{\alpha },\alpha =1,2,\cdots ,3\mathcal{N}\}$. $b_{B_{2}}(\cdots \Theta
_{\alpha }\cdots ;t)$ is the probability amplitude at moment $t$ that the
electron is in extended state $\xi _{B_{2}}$.

The e-ph interaction can cause a static displacement of atoms. The potential
energy shift caused by the displacement of atoms is:%
\begin{equation}
\Delta V=\frac{1}{2}\sum_{jk}k_{jk}x_{j}x_{k}-\sum_{p}g_{p}x_{p},
\label{pot}
\end{equation}
where $g_{p}$ is the average value of the attractive force\newline
$-\partial U(\mathbf{r},\{\mathcal{R}_{\mathbf{n}}\})/\partial X_{p}\sim
Z^{\ast}e^{2}/(4\pi\epsilon_{0}\epsilon_{s}r^{2})$ of electrons acting on
the $p^{th}$ degree of freedom in some electronic state. The second term of (%
\ref{pot}) comes from the e-ph interaction, which acts like an external
field with strength $g_{p}$.
$\Delta V$ can be written as:%
\begin{equation}
\Delta V=\frac{1}{2}\sum_{jk}k_{jk}(x_{j}-x_{j}^{0})(x_{k}-x_{k}^{0})-\frac {%
1}{2}\sum_{jk}k_{jk}x_{j}^{0}x_{k}^{0}  \label{wqs}
\end{equation}
where%
\begin{equation}
x_{m}^{0}=\sum_{p}g_{p}(k^{-1})_{mp},~m=1,2,\cdots3\mathcal{N}  \label{stad}
\end{equation}
is the static displacement for the $m^{th}$ degree of freedom. The constant
force $g_{p}$ exerted by the electron on the $p^{th}$ vibrational degree of
freedom produces a static displacement $x_{m}^{0}$ for the $m^{th}$ degree
of freedom. The deformation caused by the static external force of e-ph
interaction is balanced by the elastic force. A similar result was obtained
for a continuum model\cite{Emin76}. The last term in (\ref{wqs}) is the
polarization energy, a combined contribution from the elastic energy and
e-ph interaction.

Owing to the coupling of localized state $A_{1}$ with the vibrations of
atoms, the origin of each normal coordinate is shifted\cite{LLP}:%
\begin{equation}
\Theta _{\alpha }\rightarrow \Theta _{\alpha }-\Theta _{\alpha }^{A_{1}},~~~
\label{or}
\end{equation}%
\[
\Theta _{\alpha }^{A_{1}}=\left( M_{\alpha }\omega _{\alpha }^{2}\right)
^{-1}\sum_{p_{A_{1}}\mathbf{\in }D_{A_{1}}}d_{p_{A_{1}}}\Delta
_{p_{A_{1}}\alpha },
\]%
where $\Theta _{\alpha }^{A_{1}}\sim (N_{A_{1}}/\mathcal{N})(Z^{\ast
}e^{2}/4\pi \epsilon _{0}\epsilon _{s}\xi _{A_{1}}^{2})/\left( M_{\alpha
}\omega _{\alpha }^{2}\right) $ is the static displacement in the normal
coordinate of the $\alpha ^{th}$ mode caused by the coupling with localized
state $A_{1}$, where $N_{A_{1}}$ is the number of atoms in region $D_{A_{1}}$%
. (\ref{or}) leads to a modification of the phonon wave function and a
change in total energy. Using $(k^{-1})_{jk}=(k^{-1})_{kj}$ ~
and the inverse of (\ref{nor1}), one finds that the shift $\Theta _{\alpha
}^{A_{1}}$ of origin of the $\alpha ^{th}$ normal coordinate is related to
the static displacements by:%
\begin{equation}
\Theta _{\alpha }^{A_{1}}=\sum_{k}(\Delta ^{-1})_{\alpha
k}x_{k}^{0},~x_{k}^{0}\in D_{A_{1}}.  \label{shif}
\end{equation}%
The eigenfunctions of $h_{A_{1}}$ are:%
\[
\Psi _{A_{1}}^{\{N_{\alpha }\}}=\prod\limits_{\alpha =1}^{3\mathcal{N}}\Phi
_{N_{\alpha }}(\theta _{\alpha }-\theta _{\alpha }^{A_{1}}),~
\]%
\begin{equation}
\Phi _{N}(z)=(2^{N}N!\pi ^{1/2})^{-1/2}e^{-z^{2}/2}H_{N}(z),  \label{psia1}
\end{equation}%
where $H_{N}(z)$ is the $N^{th}$ Hermite polynomial, \newline
$\theta _{\alpha }=(M_{\alpha }\omega _{\alpha }/\hbar )^{1/2}\Theta
_{\alpha }$ is the dimensionless normal coordinate and $\theta _{\alpha
}^{A_{1}}=(M_{\alpha }\omega _{\alpha }/\hbar )^{1/2}\Theta _{\alpha
}^{A_{1}}$. The corresponding eigenvalues are:%
\begin{equation}
\mathcal{E}_{A_{1}}^{\{N_{\alpha }\}}=E_{A_{1}}^{0}+\sum_{\alpha }(N_{\alpha
}+\frac{1}{2})\hbar \omega _{\alpha }+\mathcal{E}_{A_{1}}^{b},  \label{ev}
\end{equation}%
\[
~\mathcal{E}_{A_{1}}^{b}=-\frac{1}{2}\sum_{\alpha }M_{\alpha }\omega
_{\alpha }^{2}(\Theta _{\alpha }^{A_{1}})^{2}.
\]

In an amorphous semiconductor, an electron in state $A_{1}$ polarizes the
network and the energy of state $|A_{1}\{N_{\alpha }\}\rangle $ is shifted
downward by $\mathcal{E}_{A_{1}}^{b}\sim k^{-1}[Z^{\ast }e^{2}/4\pi \epsilon
_{0}\epsilon _{s}\xi _{A_{1}}^{2}]^{2}$. The eigenvalues and eigenvectors of
$h_{B_{1}}$ are:%
\begin{equation}
\mathcal{E}_{B_{1}}^{\{N_{\alpha }\}}=E_{B_{1}}+\sum_{\alpha }(N_{\alpha }+%
\frac{1}{2})\hbar \omega _{\alpha },  \label{bv}
\end{equation}%
\[
~~~\Xi _{B_{1}}^{\{N_{\alpha }\}}=\prod\limits_{\alpha =1}^{3\mathcal{N}%
}\Phi _{N_{\alpha }}(\theta _{\alpha }).
\]

\subsection{Static displacement and vibrational amplitude}

\label{sd}

In this subsection we compare the relative magnitude of static disorder,
static displacement of atoms induced by e-ph interaction, and the amplitude
of the atomic vibrations. For a localized state $\phi_{A_{2}}$, one needs to
make following substitution in (\ref{pot}): $g_{p}=d_{p_{A_{2}}} $ ($%
d_{p_{A_{2}}}$ is defined after (\ref{ea1})) if $p\in D_{A_{2}}$ and $%
g_{p}=0 $ if $p\notin D_{A_{2}}$. The typical value of spring constant $k$
of a bond is $k\sim M\omega^{2}\sim
Z^{\ast}e^{2}/(4\pi\epsilon_{0}\epsilon_{s}d^{3})$, $M$ is the mass of a
nucleus, $\omega$ is a typical frequency of the vibrations. The static
displacement of an atom is $x_{m}^{0}\sim g/k\sim(d/\xi)^{2}d$. A typical
thermal vibrational amplitude $\mathbf{u}^{v} $ is
\[
u^{v}\sim \sqrt{k_{B}T/M\omega^{2}}\sim
d(k_{B}T)^{1/2}(Z^{\ast}e^{2}/4\pi\epsilon _{0}\epsilon_{s}d)^{-1/2}.
\]
The zero
point vibrational amplitude is
\[
\sqrt{\hbar\omega/M\omega^{2}}\sim
d(m/M)^{1/4}(\hbar^{2}/md^{2})^{1/4}(Z^{\ast}e^{2}/4\pi\epsilon_{0}%
\epsilon_{s}d)^{-1/4},
\]
where $m$ is the mass of electron. The e-ph interaction for extended states
is weak. From both experiments and simulations\cite{big,yue}, the variation
of bond length (i.e. static disorder) is of order $\sim0.05d$, where $d$ is a typical
bond length. Now it becomes clear that for \textit{amorphous semiconductors}
the static disorder is much larger than the static displacements of the
atoms induced by e-ph interaction.
The static displacement caused by the e-ph interaction is important only
when the static displacement is comparable to or larger than the amplitude of
the atomic vibrations. For weakly polar or non-polar amorphous
semiconductors, the following three statements are satisfied: (1) static
disorder localizes the carriers in band tails; (2) carriers in localized
tail states have a stronger e-ph interaction than the carriers in extended
states, and the network is polarized by the most localized tail states; and
(3) carriers in extended states have a weaker e-ph interaction and carriers
in extended states are scattered in the processes of single-phonon
absorption and emission. The small polaron theory assumed that e-ph
interaction was dominant and led to self-trapping of carriers. This
assumption is suitable for ionic crystals, molecular crystals and some polar
amorphous materials. For weakly polar or non-polar amorphous materials, the
aforementioned estimations indicate that taking carriers to be localized by
static disorder is a better starting point.
\subsection{Second quantized representation}

We expand probability amplitude $a_{A_{1}}(\cdots \Theta _{\alpha }\cdots
;t) $ with the eigenfunctions of $h_{A_{1}}$:
\begin{equation}
a_{A_{1}}=\sum_{\cdots N_{\alpha }^{\prime }\cdots }C_{\{N_{\alpha }^{\prime
}\}}^{A_{1}}(t)\Psi _{A_{1}}^{\{N_{\alpha }^{\prime }\}}e^{-it\mathcal{E}%
_{A_{1}}^{\{N_{\alpha }^{\prime }\}}/\hbar },  \label{az}
\end{equation}%
where $C_{\{N_{\alpha }^{\prime }\}}^{A_{1}}(t)$ is the probability
amplitude at moment $t$ that the electron is in localized state $A_{1}$
while the vibrational state of the nuclei is characterized by occupation
number $\{N_{\alpha }^{\prime },\alpha =1,2,\cdots ,3\mathcal{N}\}$.
Similarly we expand the probability amplitude $b_{B_{1}}(\cdots \Theta
_{\alpha }\cdots ;t)$ with eigenfunctions of $h_{B_{1}}$:%
\begin{equation}
b_{B_{1}}=\sum_{\cdots N_{\alpha }^{\prime }\cdots }F_{\{N_{\alpha }^{\prime
}\}}^{B_{1}}(t)\Xi _{B_{1}}^{\{N_{\alpha }^{\prime }\}}e^{-it\mathcal{E}%
_{B_{1}}^{\{N_{\alpha }^{\prime }\}}/\hbar },  \label{bz}
\end{equation}%
where $F_{\{N_{\alpha }^{\prime }\}}^{B_{1}}(t)$ is the probability
amplitude at moment $t$ that the electron is in extended state $B_{1}$ while
the vibrational state of the nuclei is characterized by occupation number $%
\{N_{\alpha }^{\prime },\alpha =1,2,\cdots ,3\mathcal{N}\}$.

Substitute Eq.(\ref{az}) and Eq.(\ref{bz}) into Eq.(\ref{1s}) and applying $%
\int\displaystyle\prod
\limits_{\alpha}d\theta_{\alpha}\Psi_{A_{2}}^{\{N_{\alpha}\}}$ to both sides
we obtain

\[
i\hbar \frac{\partial C_{\{N_{\alpha }\}}^{A_{2}}(t)}{\partial t}%
=\sum_{A_{1}\cdots N_{\alpha }^{\prime }\cdots }\langle A_{2}\{N_{\alpha
}\}|V_{LL}^{tr}|A_{1}\{N_{\alpha }^{\prime }\}\rangle
\]%
\[
C_{\{N_{\alpha }^{\prime }\}}^{A_{1}}(t)e^{it(\mathcal{E}_{A_{2}}^{\{N_{%
\alpha }\}}-\mathcal{E}_{A_{1}}^{\{N_{\alpha }^{\prime }\}})/\hbar }
\]%
\[
+\sum_{B_{1}\cdots N_{\alpha }^{\prime }\cdots }\langle A_{2}\{N_{\alpha
}\}|V_{EL}^{e-ph}|B_{1}\{N_{\alpha }^{\prime }\}\rangle
\]%
\begin{equation}
F_{\{N_{\alpha }^{\prime }\}}^{B_{1}}(t)e^{it(\mathcal{E}_{A_{2}}^{\{N_{%
\alpha }\}}-\mathcal{E}_{B_{1}}^{\{N_{\alpha }^{\prime }\}})/\hbar }
\label{cq}
\end{equation}%
where%
\[
\langle A_{2}\{N_{\alpha }\}|V_{LL}^{tr}|A_{1}\{N_{\alpha }^{\prime
}\}\rangle =
\]%
\begin{equation}
J_{A_{2}A_{1}}\int \displaystyle\prod\limits_{\alpha }d\theta _{\alpha }\Psi
_{A_{2}}^{\{N_{\alpha }\}}\Psi _{A_{1}}^{\{N_{\alpha }^{\prime }\}}
\label{LLT}
\end{equation}%
describes the transition from localized state $A_{1}$ with phonon
distribution $\{\cdots N_{\alpha }^{\prime }\cdots \}$ to localized state $%
A_{2}$ with phonon distribution $\{\cdots N_{\alpha }\cdots \}$ caused by
transfer integral $J_{A_{2}A_{1}}$ defined in Eq.(\ref{tra1}).%
\[
\langle A_{2}\{N_{\alpha }\}|V_{EL}^{e-ph}|B_{1}\{N_{\alpha }^{\prime
}\}\rangle =
\]%
\begin{equation}
\int \displaystyle\prod\limits_{\alpha }d\theta _{\alpha }\Psi
_{A_{2}}^{\{N_{\alpha }\}}(\sum_{\alpha }\Theta _{\alpha
}K_{A_{2}B_{1}}^{\prime \alpha })\Xi _{B_{1}}^{\{N_{\alpha }^{\prime }\}}
\label{ELT}
\end{equation}%
is the transition from an extended state to a localized state induced by
electron-phonon interaction.

Similarly from Eq.(\ref{2s}) we have%
\[
i\hbar \frac{\partial F_{\{N_{\alpha }\}}^{B_{2}}}{\partial t}%
=\sum_{A_{1}\cdots N_{\alpha }^{\prime }\cdots }\langle B_{2}\{N_{\alpha
}\}|V_{LE}^{tr}|A_{1}\{N_{\alpha }^{\prime }\}\rangle
\]%
\[
C_{\{N_{\alpha }^{\prime }\}}^{A_{1}}e^{it(\mathcal{E}_{B_{2}}^{\{N_{\alpha
}\}}-\mathcal{E}_{A_{1}}^{\{N_{\alpha }^{\prime }\}})/\hbar }
\]%
\[
+\sum_{B_{1}\cdots N_{\alpha }^{\prime }\cdots }\langle B_{2}\{N_{\alpha
}\}|V_{EE}^{e-ph}|B_{1}\{N_{\alpha }^{\prime }\}\rangle
\]%
\begin{equation}
F_{\{N_{\alpha }^{\prime }\}}^{B_{1}}e^{it(\mathcal{E}_{B_{2}}^{\{N_{\alpha
}\}}-\mathcal{E}_{B_{1}}^{\{N_{\alpha }^{\prime }\}})/\hbar }  \label{fq}
\end{equation}%
where%
\[
\langle B_{2}\{N_{\alpha }\}|V_{LE}^{tr}|A_{1}\{N_{\alpha }^{\prime
}\}\rangle =
\]%
\begin{equation}
J_{B_{2}A_{1}}^{\prime }\int \displaystyle\prod\limits_{\alpha }d\theta
_{\alpha }\Xi _{B_{2}}^{\{N_{\alpha }\}}\Psi _{A_{1}}^{\{N_{\alpha }^{\prime
}\}}  \label{LET}
\end{equation}%
describes the transition from localized state $|A_{1}\cdots N_{\alpha
}^{\prime }\cdots \rangle $ to extended state $|B_{2}\cdots N_{\alpha
}\cdots \rangle $ caused by transfer integral $J_{B_{2}A_{1}}^{\prime }$,
the dependence on $\{x_{j}\}$ in $J^{\prime }$ is neglected.%
\[
\langle B_{2}\{N_{\alpha }\}|V_{EE}^{e-ph}|B_{1}\{N_{\alpha }^{\prime
}\}\rangle =
\]
\begin{equation}
\int \displaystyle\prod\limits_{\alpha }d\theta _{\alpha }\Xi
_{B_{2}}^{\{N_{\alpha }\}}(\sum_{\alpha }\Theta _{\alpha
}K_{B_{2}B_{1}}^{\alpha })\Xi _{B_{1}}^{\{N_{\alpha }^{\prime }\}}
\label{ee1}
\end{equation}%
is the matrix element of the transition between two extended states caused
by electron-phonon interaction.
Eq.(\ref{cq}) and Eq.(\ref{fq}) are the evolution equations in
second-quantized form. The phonon state on the left hand side (LHS) can be
different from that in the right hand side. In general, the occupation
number in each mode changes when the electron changes its state.

\section{Transition between two localized states}

\label{LL}

\subsection{$J_{_{A_{3}A_{1}}}$ as perturbation}

\label{JP} In amorphous solids, the transfer integral (\ref{tra1}) between
two localized states is small. Perturbation theory can be used to solve Eq.(%
\ref{cq}) to find the probability amplitude. Then the transition probability%
\cite{Hol2} from state $\Psi _{\{N_{\alpha }^{\prime }\}}^{A_{1}}$ to state $%
\Psi _{\{N_{\alpha }\}}^{A_{3}}$ is:
\[
W_{T}(A_{1}\rightarrow A_{3})=\frac{J_{A_{3}A_{1}}^{2}}{\hbar ^{2}}\exp \{%
\frac{-\beta }{2}[(E_{A_{3}}^{0}+\mathcal{E}_{b}^{A_{3}})-(E_{A_{1}}^{0}+%
\mathcal{E}_{b}^{A_{1}})]\}\times
\]%
\[
\exp \{-\frac{1}{2}\sum_{\alpha }(\theta _{\alpha }^{A_{3}}-\theta _{\alpha
}^{A_{1}})^{2}\coth \frac{^{\beta \hbar \omega _{\alpha }}}{2}\}\times
\]%
\[
\int_{-t}^{t}d\tau \exp \{\frac{i\tau }{\hbar }[(E_{A_{3}}^{0}+\mathcal{E}%
_{b}^{A_{3}})-(E_{A_{1}}^{0}+\mathcal{E}_{b}^{A_{1}})]\}
\]%
\begin{equation}
\lbrack \exp \{\frac{1}{2}\sum_{\alpha }(\theta _{\alpha }^{A_{3}}-\theta
_{\alpha }^{A_{1}})^{2}csch\frac{\beta \hbar \omega _{\alpha }}{2}\cos \tau
\omega _{\alpha })\}-1].  \label{TT}
\end{equation}%
We should notice: (i) for localized states we adopt the partition (%
\ref{sl}) for full potential energy (cf. Appendix\ref{der}), the LL
transition is driven by transfer integral $J_{A_{3}A_{1}}$; (ii) The
localized band tail states strongly couple with the atomic vibrations\cite%
{dra,tafn}, a carrier in a localized tail state introduces static
displacements of atoms through e-ph interaction, so that the occupied
localized state couples with all vibrational modes. When a carrier moves in
or out of a localized tail state, the atoms close to this state are shifted.
In normal coordinate language, this is expressed by $\theta _{\alpha
}^{A_{3}}-\theta _{\alpha }^{A_{1}}$ in (\ref{TT}) for each mode. Thus a LL
transition is a multi-phonon process; (iii) If we notice that the transfer
integral $J_{A_{3}A_{1}}\propto e^{-R_{31}/\xi }$, where $2/\xi =\xi
_{A_{1}}^{-1}+\xi _{A_{3}}^{-1}$. The product of first two factors in (\ref%
{TT}) is similar to the single-phonon transition probability obtained in
\cite{mil}. In the following subsection, we will see that (\ref{TT}) is
reduced to MA theory when reorganization energy is small or two localized
states are similar: $\theta _{\alpha }^{A_{3}}\thickapprox \theta _{\alpha
}^{A_{1}}$ or when temperature is high.

\subsection{High temperature limit}

\label{ht} For high temperature ($k_{B}T\geq \hbar \overline{\omega }$, $%
\overline{\omega }=2\pi \bar{\nu}$), (\ref{TT}) reduces to:%
\[
W_{T}(A_{1}\rightarrow A_{3})=\frac{J_{A_{3}A_{1}}^{2}}{\hbar ^{2}}\exp \{%
\frac{-\beta }{2}[(E_{A_{3}}^{0}+\mathcal{E}_{b}^{A_{3}})-(E_{A_{1}}^{0}+%
\mathcal{E}_{b}^{A_{1}})]\}\times
\]%
\[
\exp \{-\frac{1}{2}\sum_{\alpha }(\theta _{\alpha }^{A_{3}}-\theta _{\alpha
}^{A_{1}})^{2}\tanh \frac{{\beta \hbar \omega _{\alpha }}}{4}\}\times
\]%
\[
(2\pi )^{1/2}[\frac{1}{2}\sum_{\alpha }(\theta _{\alpha }^{A_{3}}-\theta
_{\alpha }^{A_{1}})^{2}\omega _{\alpha }^{2}csch\frac{\beta \hbar \omega
_{\alpha }}{2}]^{-1/2}
\]%
\begin{equation}
\exp (-\frac{[(E_{A_{3}}^{0}+\mathcal{E}_{b}^{A_{3}})-(E_{A_{1}}^{0}+%
\mathcal{E}_{b}^{A_{1}})]^{2}}{\sum_{\alpha }(\theta _{\alpha
}^{A_{3}}-\theta _{\alpha }^{A_{1}})^{2}\hbar ^{2}\omega _{\alpha }^{2}csch%
\frac{\beta \hbar \omega _{\alpha }}{2}}).  \label{go}
\end{equation}%
%
%
%
%
%
%

At `very' high temperature ($k_{B}T\geq 2.5\hbar \overline{\omega }$) using $%
\tanh x\approx x$ and csch$x\approx 1/x$, (\ref{go}) becomes%
\[
W_{T}(A_{1}\rightarrow A_{3})=\nu _{LL}e^{-E_{a}^{LL}/k_{B}T},
\]%
\begin{equation}
~\nu _{LL}=\frac{J_{A_{3}A_{1}}^{2}}{\hbar }[\frac{\pi }{\lambda _{LL}k_{B}T}%
]^{1/2},~~E_{a}^{LL}=\frac{\lambda _{LL}}{4}(1+\frac{\Delta G_{LL}^{0}}{%
\lambda _{LL}})^{2}.  \label{Mar}
\end{equation}%
where%
\[
\Delta G_{LL}^{0}=(E_{A_{3}}^{0}+\mathcal{E}_{b}^{A_{3}})-(E_{A_{1}}^{0}+%
\mathcal{E}_{b}^{A_{1}}),
\]%
\begin{equation}
\lambda _{LL}=\frac{1}{2}\sum_{\alpha }M_{\alpha }\omega _{\alpha
}^{2}(\Theta _{\alpha }^{A_{3}}-\Theta _{\alpha }^{A_{1}})^{2}.  \label{de}
\end{equation}

$\Delta G_{LL}^{0}$ is the energy difference between two localized states. $%
\lambda_{LL}$ is the reorganization energy which depends on the vibrational
configurations $\{\Theta_{\alpha}^{A_{3}}\}$ and $\{\Theta_{\alpha}^{A_{1}}%
\} $ of the two localized states. Because $\Theta_{\alpha}^{A}$ does not
have a determined sign for different states and modes, one can only roughly
estimate
\[
\lambda_{LL}\sim k^{-1}(Z^{\ast}e^{2}/4\pi\epsilon_{s}\epsilon_{0}\xi
^{2})^{2}\sim\epsilon_{s}^{-1}(d/\xi)^{3}(Z^{\ast}e^{2}/4\pi\epsilon_{0}%
\xi).
\]
This is consistent with common experience: the longer the localization
length (the weaker the localization), the smaller the reorganization energy.
From (\ref{Mar}), we know that $E_{a}^{LL}$ is about 0.01-0.05eV, in
agreement with the observed value\cite{tes} for a-Si. (\ref{Mar}) has the
same form as Marcus type rate (\ref{ea}) for electron transfer in a polar
solvent and in large molecules. Because $x_{0}\ge u^{v}$, the vibrational
energy $ku^{v2}/2$ is the lower limit of the reorganization energy $%
\lambda_{LL}\sim N_{A}kx_{0}^{2}/2$.
For most LL transitions, $\lambda_{LL}$ is greater than $\Delta G_{LL}$. For
less localized states and higher temperature, $\lambda_{LL}\sim\Delta G_{LL}$%
, then $E_{a}^{LL}=\lambda/4+\Delta G_{LL}/2+(\Delta
G_{LL})^{2}/4\lambda\simeq\Delta G_{LL}$, and the present work reduces to MA
theory.

From \textit{ab initio} simulations\cite{big,yue} in various a-Si structural
models, the distance between the two nearest most localized tail states is $%
R_{A_{3}A_{1}}\sim 3-5$\AA ~(one or two bond lengths). The effective nuclear
charge\cite{cle} is $Z^{\ast}=4.29$ and static dielectric constant\cite{si} $%
\epsilon_{s}=11.8$, $J_{A_{3}A_{1}}\sim0.02$eV (Appendix B). The energy
dependence $\Delta G_{LL}$ between the final and initial states affects $%
W_{LL}$, mobility and the contribution to conductivity. For LL transition, the
largest $\Delta G_{LL}$ is the mobility edge $D$, so that we pick up $%
D/2=0.05$eV\cite{sca} as a typical $\Delta G_{LL}$. For the most localized
tail state in a-Si, the localization length\cite{big,yue} is $\xi\approx 5$%
\AA . $J_{A_{1}A_{2}}$ is estimated in Appendix B. From the force constant%
\cite{force} $k\sim dc_{44},$ $c_{44}=81$ GPa, a typical reorganization
energy is $\lambda_{LL}=0.2$eV, yielding $W_{T}\sim10^{12}$sec$^{-1}$.

If one assumes the same parameters as above, the prediction of MA theory
would be \newline
$(n_{D/2}$~ or~ $n_{D/2}+1)J^{2}/\hbar(D/2)\sim10^{12}-10^{13}$sec $^{-1}$
(at T=300K), the same order of magnitude as the present work, where $%
n_{D/2}=(e^{\beta D/2}-1)^{-1}$ is the phonon occupation factor. This is why
MA theory appears to work for higher temperature.

\subsection{Field dependence of conductivity}

\label{fd}

For electrons, external field $F$ lowers the barrier of the LL
transition $\delta (\Delta G_{LL})=-eF\xi -eFR<0$ along the direction
opposite to the field, where $R$ is the distance between centers of two
localized states.

An electric field increases the localization length of a localized state.
A localized electron is bound by the extra force $f\sim Z^{\ast
}e^{2}u^{s}/4\pi \epsilon _{0}\epsilon _{s}d^{3}$ of the disorder potential:
$fu^{s}=\hbar ^{2}/2m\xi ^{2}$. The relative change $\delta \xi $ in
localization length induced by the external field is $\delta \xi /\xi
=-\delta f/2f,$ where $\delta f=-eF$ is the force exerted by external
electric field. Thus $\delta \xi /\xi \thicksim (eF/2)(Z^{\ast
}e^{2}u^{s}/[4\pi \epsilon _{0}\epsilon _{s}d^{3}])^{-1}>0$. As a
consequence, reorganization energy $\lambda $ decreases with increasing $F$.
From $\lambda \thicksim g^{2}/k\thicksim \epsilon _{s}^{-2}Z^{\ast
}e^{2}d^{3}/(4\pi \epsilon _{0}\xi ^{4})$, the relative change $\delta
\lambda $ in reorganization energy $\lambda $ is $\delta \lambda /\lambda
=-4\delta \xi /\xi \thickapprox -2eF(Z^{\ast }e^{2}u^{s}/[4\pi \epsilon
_{0}\epsilon _{s}d^{3}])^{-1}<0$.

From the expression of $E_{LL}^{a}$ in (\ref{Mar}), to first order of field,
the change in activation energy is%
\begin{equation}
\delta E_{LL}^{a}=\frac{\delta \lambda }{4}[1-(\frac{\Delta G}{\lambda }%
)^{2}]+\frac{\delta (\Delta G)}{2}(1+\frac{\Delta G}{\lambda }).  \label{chj}
\end{equation}%
For temperatures lower than the Debye temperature, $\Delta G_{LL}<\lambda _{LL}$.
It is obvious that $\delta E_{LL}^{a}<0$, activation energy decreases with
external field.

Increasing $\xi $ with $F$ also leads to that transfer integral $J$
increases with $F$: since $J_{A_{3}A_{1}}\propto e^{-R_{31}/\xi }$, \\$%
J_{A_{3}A_{1}}(F)/J_{A_{3}A_{1}}(0)=\exp \{R_{31}(\xi ^{-1}-\xi
_{F}^{-1})\} \\ \thickapprox \exp \{\xi ^{-1}R_{31}\delta \xi /\xi \}>1$, where $%
\xi _{F}$ is the average localization length in external field. Using the
value of $\delta \xi /\xi $, $J_{A_{3}A_{1}}(F)/J_{A_{3}A_{1}}(0)\\=\exp \{\xi
_{0}^{-1}R_{31}(eF/2)(Z^{\ast }e^{2}u^{s}/[4\pi \epsilon _{0}\epsilon
_{s}d^{3}])^{-1}\}$.

For hopping conduction, the conductivity $\sigma $ is estimated as $\sigma
=ne^{2}\mu $, for $n$ carrier density and $\mu =D/k_{B}T$ the mobility, $D$ is
the diffusion coefficient of carriers\cite{md}. To obtain conductivity, one
should average mobility over different $\Delta G_{LL}$, density of states
and occupation number. We approximate this average by $\Delta G_{LL}\sim
k_{B}T$. If one only considers the contribution from the hopping among
nearest neighbor localized states, $D=R_{A_{3}A_{1}}^{2}W_{T}(A_{1}%
\rightarrow A_{3})$. The force produced by the experimental field is much weaker than the extra force produced by the static disorder $eF<<f$, no carrier is
delocalized by the external field. The carrier density $n$ and the distance $R_{A_{3}A_{1}}$
between two localized states are not affected by external field, so that $%
\sigma (T,F)/\sigma (T,0)=W_{T}(F)/W_{T}(0)$. According to (\ref{Mar}), $%
\sigma (T,F)/\sigma (T,0)=[\lambda (F)/\lambda
(0)]^{-1/2}[J_{A_{3}A_{1}}(F)/J_{A_{3}A_{1}}(0)]^{2}\\ \exp \{-\beta \lbrack
E_{a}^{LL}(F)-E_{a}^{LL}(0)]\}$. Workers often fit experimental data in form:
$\sigma (T,F)/\sigma (T,0)=\exp [s(T)F]$. Using $(1+x)^{-1/2}\thickapprox
1-x/2\thickapprox e^{-x/2}$ transform $(1+\delta \lambda /\lambda
)^{-1/2}\thickapprox e^{-\delta \lambda /(2\lambda )}$, one finds:%
\[
s(T)=e(\frac{Z^{\ast }e^{2}u^{s}}{4\pi \epsilon _{0}\epsilon _{s}d^{3}(T)}%
)^{-1}+\frac{R}{\xi (T)}e(\frac{Z^{\ast }e^{2}u^{s}}{4\pi \epsilon
_{0}\epsilon _{s}d^{3}(T)})^{-1}
\]%
\[
+\frac{\lambda (T)}{4k_{B}T}(1-\frac{\Delta G_{LL}^{2}}{\lambda ^{2}(T)})2e(%
\frac{Z^{\ast }e^{2}u^{s}}{4\pi \epsilon _{0}\epsilon _{s}d^{3}(T)})^{-1}
\]%
\begin{equation}
+\frac{e(\xi (T)+R)}{2k_{B}T}(1+\frac{\Delta G_{LL}}{\lambda (T)}).
\label{fcon}
\end{equation}

According to the percolation theory of the localized-delocalized transition%
\cite{zal}, the localization length $\xi$ of a localized state increases
with rising temperature: $\xi(T)=\xi_{0}(1-T/T_{m})^{-1}$ (the critical
index is between 1/2 and 1; we employ 1 here), where $T_{m}$ is the
temperature where all localized states become delocalized; $T_{m}$ is close
to the melting point.
Then $\lambda(T)=g^{2}(T)/k(T)=\lambda_{0}(1-T/T_{m})^{4}$ and the slope $%
s(T)$ in $\exp[s(T)F]$ increases with decreasing temperature. Figure \ref%
{aGe} is a comparison between the observations\cite{ell} in a-Ge and the
values of present work. The parameters used are 
$d=2.49$\AA , $u^{s}/d=0.1$ and $T_{m}=$1210 K. $\lambda_{0}=0.2$eV is
estimated from $Z^{\ast}=4$. Because the conductivity comes from various
localized states, $\Delta G_{LL}$ varies from 0 to $D$.


The field polarizes the wave functions of occupied states and empty states
(with a virtual positive charge). A static voltage on a sample adds a term
to the double-well potential between two localized states:%
\begin{equation}
U(y)=\frac{1}{2}ay^{2}+\frac{1}{4}by^{4}-eFy,  \label{fe1}
\end{equation}
where $a\sim-k$ and $b\sim k/x_{0}^{2}$. To first order in the field, the
two minima $y_{1}$ and $y_{2}$ of (\ref{fe1}) do not shift. To second order
in field, the distance between two minima of (\ref{fe1}) decreases by an
amount $\delta R=(3b/4a^{3})\sqrt{-a/b}(eF)^{2}$. This results in a further
decrease of reorganization energy\cite{marcu} in addition to the direct
voltage drop.

\begin{figure}[tbp]
\resizebox{0.5\textwidth}{!}{  \includegraphics{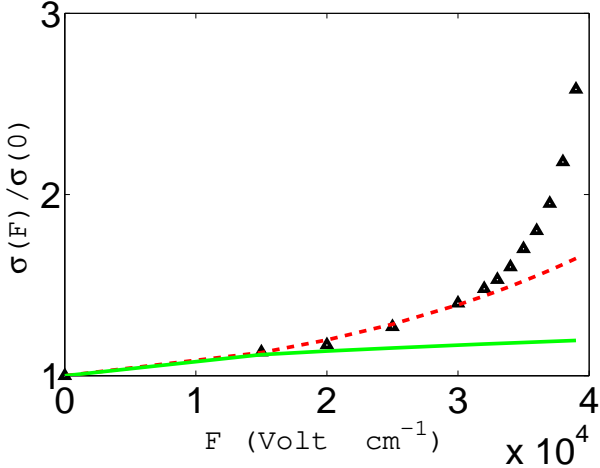}
} 
\caption{Field dependence of conductivity in vanadium oxide. The triangle
symbols are measured \protect\cite{rog} values of $\protect\sigma(F)/\protect%
\sigma(F=0)$ for VO$_{1.83}$ at 200K. The solid line is the change in
carrier density $n(E)/n(E=0)$ computed from Frenkel-Poole model, dashed line
is the change in mobility $\protect\mu(F)/\protect\mu(F=0)$ from present
work.}
\label{Frenkel}
\end{figure}

In Figure~ \ref{Frenkel}, we compare the observed $\sigma(F)/\sigma(F=0)$
results\cite{rog} at 200K with the best fit of $\exp($const$\cdot F^{1/2})$
of the Frenkel-Poole model and with $\exp(s(T)F+dF^{2})$ of present work (VO$%
_{1.83}$ is of special interest for microbolometer applications\cite{rog}),
where
\[
d=(k_{B}T)^{-1}(Z^{\ast}e^{2}/4\pi\epsilon_{0}\epsilon_{s}R^{2})(3b/4a^{3})%
\sqrt{-a/b}e^{2}.
\]
We can see from Figure~\ref{Frenkel} that the change in mobility provides a
better description of experiments than that of the change in carrier density
by field in the Frenkel-Poole model\cite{fren}.

\subsection{Meyer-Neldel rule}

\label{mnr1} From the formula for $\lambda_{LL}$ in the paragraph below (\ref{fcon}), the reorganization energy for LL transitions decreases with
rising temperature:
\[
\delta\lambda_{LL}(T)=-4\delta T(1-T/T_{m})^{-1}T_{m}^{-1}\lambda_{LL}(T).
\]
From (\ref{Mar}),
\begin{equation}
\delta E_{a}^{LL}(T)=-E_{a}(0)\frac{\delta T}{T_{m}}(1-\frac{T}{T_{m}}%
)^{-1}(1-\frac{\Delta G_{LL}^{2}}{\lambda_{LL}^{2}})\frac{\lambda_{LL}}{%
E_{a}^{LL}(0)}.  \label{cha}
\end{equation}
If $E_{a}^{LL}(T)$ decreases with rising temperature according to:
\begin{equation}
E_{a}^{LL}(T)=E_{a}^{LL}(0)(1-T/T_{MN}),  \label{mnr}
\end{equation}
then the Meyer-Neldel rule is obtained\cite{abt}: comparing (\ref{cha}) and (%
\ref{mnr}) one finds%
\begin{equation}
T_{MN}\approx\frac{T_{m}}{4}(1+\frac{\Delta G_{LL}}{\lambda_{LL}})(1-\frac{%
\Delta G_{LL}}{\lambda_{LL}})^{-1}.  \label{mnt}
\end{equation}
According to (\ref{go}), during a LL transition the vibrational
configurations of two localized tail states are reorganized. A large number
of excitations (phonons) is required. A temperature-dependent activation
energy implies that {\textit{entropy}} must be involved and is an important
ingredient for activation. The present approach supports the
multi-excitation entropy theory of Yelon and Movaghar\cite{Yelon90,Yelon06}.

\begin{center}
\begin{table*}[ht]
\caption{predicted and observed Meyer-Neldel temperature T$_{MN}$ in several
materials (see text).}
\label{tab:tmn}{\small \hfill{}
\begin{tabular}{cccccccc}
\hline\hline
material & T$_{m}$(K) & $\epsilon_{s}$ & d{(\AA )} & $\xi/d$ & $\Delta
G_{LL} $(eV) & T$_{MN}^{theory}$(K) & T$_{MN}^{expt}$(K) \\ \hline
a-Si:H & 1688 & 11.9 & 2.35 & 1.7-1.8 & 0.1 & 756-890 & 499-776\cite{Over}
\\ \hline
a-Ge$_{1-x}$Se$_{2}$Pb$_{x}$ & 993 & 13 & 2.41 & 1.7 & 0.11 & 744 & 765\cite%
{eln} \\ \hline
a-(As$_{2}$Se$_{3}$)$_{100-x}$(SbSI)$_{x}$ & 650 & 11 & 2.4 & 1.5 & 0.12 &
596 & 591\cite{sku} \\ \hline
ZnO & 2242 & 9.9-11 & 1.71 & 1.5 & 0.01 & 595 & 226-480\cite{schm,sag} \\
\hline
NiO & 2257\cite{winch} & 11\cite{moni} & 1.75 & 1 & 0.06\cite{moni,elp} &
1455 & 1460-1540\cite{dew} \\ \hline\hline
\end{tabular}%
}
\end{table*}
\end{center}

Table \ref{tab:tmn} is a comparison of the predicted Meyer-Neldel
temperature $T_{MN}$ with observed ones. The number of atoms involved in a
localized tail state is taken to the second nearest neighbor. The
reorganization energy is estimated from the parameters given in Table \ref%
{tab:tmn}, then T$_{MN}$ is estimated from (\ref{mnt}). Beside NiO, the most
localized state extends to about $1.5d$. In NiO the hole of d electron shell
is localized on one oxygen atom. The localization comes from the on-site
repulsion in a d-band split by the crystal field. The theory agrees well
with observations in quite different materials. In a typical ionic crystal
like ZnO, the localized tail states arise from thermal disorder and are
confined in very small energy range: $E_{U}\sim ku^{v2}/2\sim
k_{B}T\sim0.025 $eV (T=300K). The fraction of localized carriers is much
less than that in an amorphous semiconductor where localization is caused by
static disorder. That is why $\Delta G_{LL}$ is about 10 times smaller than
that in an amorphous semiconductor. In ZnO, most of the carriers are better
described by large polarons. Taking the melting point as the
localized-delocalized transition temperature $T_{m}$ is presumably an
overestimation, so that the computed T$_{MN}$ is too high.

\subsection{Low temperatures}

\label{Lt} For low temperature ($k_{B}T\leq \hbar \bar{\omega}/10$),
the argument in the last exponential of (\ref{TT}) is small. The exponential
can be expanded in Taylor series and the `time' integral can be completed.
Denote:%
\begin{equation}
f(\omega _{\alpha })=\frac{1}{2}(\theta _{\alpha }^{A_{3}}-\theta _{\alpha
}^{A_{1}})^{2}csch\frac{\beta \hbar \omega _{\alpha }}{2},  \label{f}
\end{equation}%
then:%
\[
W_{T}(A_{1}\rightarrow A_{3})=\frac{2\pi J_{A_{3}A_{1}}^{2}}{\hbar }\exp \{%
\frac{-\beta \Delta G_{LL}^{0}}{2}\}
\]%
\[
\exp \{-\frac{1}{2}\sum_{\alpha }(\theta _{\alpha }^{A_{3}}-\theta _{\alpha
}^{A_{1}})^{2}\coth \frac{^{\beta \hbar \omega _{\alpha }}}{2}\}
\]%
\[
\{\sum_{\alpha }f(\omega _{\alpha })\frac{1}{2}[\delta (\Delta
G_{LL}^{0}+\hbar \omega _{\alpha })+\delta (\Delta G_{LL}^{0}-\hbar \omega
_{\alpha })]
\]%
\[
+\sum_{\alpha \alpha ^{\prime }}f(\omega _{\alpha })f(\omega _{\alpha
^{\prime }})\frac{1}{8}[\delta (\Delta G_{LL}^{0}+\hbar \omega _{\alpha
}+\hbar \omega _{\alpha ^{\prime }})
\]

\[
+\delta(\Delta
G_{LL}^{0}-\hbar\omega_{\alpha}-\hbar\omega_{\alpha^{\prime}})
\]

\begin{equation}
+\delta (\Delta G_{LL}^{0}+\hbar \omega _{\alpha }-\hbar \omega _{\alpha
^{\prime }})+\delta (\Delta G_{LL}^{0}-\hbar \omega _{\alpha }+\hbar \omega
_{\alpha ^{\prime }})]\newline
+\cdots \}.  \label{di}
\end{equation}%
One may say that the transfer integral is reduced by a factor $\exp \{-\frac{%
1}{4}\sum_{\alpha }(\theta _{\alpha }^{A_{3}}-\theta _{\alpha
}^{A_{1}})^{2}\}$ due to the strong e-h coupling of localized states.

\begin{figure}[tbp]
\resizebox{0.5\textwidth}{!}{  \includegraphics{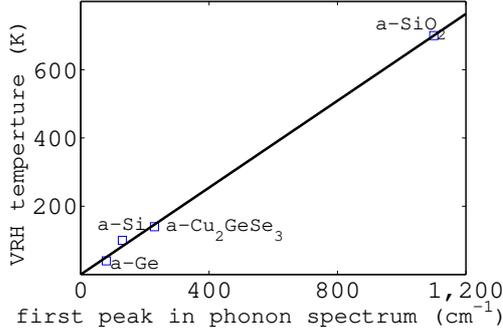}
} 
\caption{The present work predicts that the upper temperature limit of VRH
is proportional to $\bar{\protect\nu}$. The squares are
observed data in several materials\protect\cite%
{brod74,bah,sriv,awa,oss,gmar,aiy}, the line is a linear fit.}
\label{peak}
\end{figure}

The derivation of (\ref{di}) from (\ref{TT}) suggests that below a certain temperature T$_{up}$, the thermal vibrations of network do not have enough energy to adjust the atomic static displacements around localized states. The low temperature LL transition (\ref{di}) becomes the only way to cause a transitions between two localized states (it does not need reorganization energy). Since the variable range hopping (VRH)\cite{motv} is the most probable low temperature LL transition, T$_{up}$ is the upper limit temperature of VRH.

On the other hand, the available thermal energy is $\sum_{j}\hbar\omega_{j}n_{j}$, where $n_{j}$ is the occupation number of the $j^{th}$ mode. At a low temperature,  $n_{j}\sim exp(-\hbar\omega_{j}/k_{B}T)$, the available vibrational energy is determined by the number of modes with $\hbar\omega\le k_{B}$T. The higher the frequency $\bar{\nu}$ of the first peak in phonon spectrum, the fewer the excited phonons. In other words, for a system with higher $\bar{\nu}$, only at higher temperature could one have enough vibrational energy to enable a transition. 
Therefore the present work predicts that for different materials, the upper temperature limit $T_{up}$ of VRH is proportional to the frequency $\bar{\nu}$ of the first peak in phonon spectrum.

Figure \ref{peak} reports experimental data for a-Si\cite{brod74}, a-Ge\cite{brod74,bah}, a-SiO$_{2}$\cite{sriv,awa,oss} and a-Cu$_{2}$GeSe$_{3}$\cite{gmar,aiy}: the upper limit temperature T$_{up}$ of VRH vs. the first peak of phonon spectrum $\bar{\nu}$. A linear relation between T$_{up}$ and $\bar{\nu}$ is satisfied. From a linear fit, we deduce $T_{up}\approx\hbar\overline{\omega}/2.3k_{B}$, far beyond the more stringent condition T$<\hbar\overline{\omega}/10k_{B}$ for Taylor expansion of the exponential in (\ref{TT}). This is easy to understand: low frequency modes are acoustic, the density of states decreases quickly with reducing phonon frequency (Debye square distribution). T$<\hbar\overline{\omega}/10k_{B}$ is derived from csch$\frac{\beta\hbar\omega}{2}<0.01$ for all modes. The exponent of (\ref{TT}) is a summation over all modes. At $T_{up}\approx\hbar\overline{\omega}/2.3k_{B}$, although a single phonon seems have higher frequency, because the density of states at this frequency is small, the available vibration energy is low and VRH is already dominant.

\section{Transition from a localized state to an extended state}

\label{LE} The transition probability from localized state $A_{1}$ to
extended state $B_{2}$ is:%
\[
W_{T}(A_{1}\rightarrow B_{2})=\frac{J_{B_{2}A_{1}}^{\prime 2}}{\hbar ^{2}}%
e^{-\beta (E_{B_{2}}-E_{A_{1}}^{0}-\mathcal{E}_{A_{1}}^{b})/2}
\]%
\[
\exp \{-\frac{1}{2}\sum_{\alpha }(\theta _{\alpha }^{A_{1}})^{2}\coth \frac{%
\beta \hbar \omega _{\alpha }}{2}\}\times
\]%
\[
\int_{-t}^{t}d\tau \exp \{\frac{i\tau }{\hbar }(E_{B_{2}}-E_{A_{1}}^{0}-%
\mathcal{E}_{A_{1}}^{b})\}
\]%
\begin{equation}
\lbrack \exp \{\frac{1}{2}\sum_{\alpha }(\theta _{\alpha }^{A_{1}})^{2}csch%
\frac{\beta \hbar \omega _{\alpha }}{2}\cos \omega _{\alpha }\tau \}-1].
\label{ple4}
\end{equation}%
When a carrier moves out of a localized state, the nearby atoms are shifted
and the occupation number in all modes are changed. Thus a LE transition is a multi-phonon process.

For `very' high temperature $k_{B}T\geq 2.5\hbar\overline{\omega},$ (\ref%
{ple4}) reduces to:%
\begin{equation}
W_{T}^{LE}=\nu_{LE}e^{-E_{a}^{LE}/k_{B}T},~~E_{a}^{LE}=\frac{\lambda_{LE}}{4}%
(1+\frac{\Delta G_{LE}^{0}}{\lambda_{LE}})^{2},  \label{LEmar}
\end{equation}
where%
\begin{equation}
\Delta G_{LE}^{0}=E_{B_{2}}-(E_{A_{1}}^{0}+\mathcal{E}_{A_{1}}^{b})
\label{ELd}
\end{equation}
is the energy difference between extended state $B_{2}$ and localized state $%
A_{1}$, and
\begin{equation}
\nu_{LE}=\frac{J_{B_{2}A_{1}}^{\prime2}}{\hbar}(\frac{\pi}{\lambda_{LE}k_{B}T%
})^{1/2},~\lambda_{LE}=\frac{1}{2}\sum_{\alpha}M_{\alpha}\omega_{%
\alpha}^{2}(\Theta_{\alpha}^{A_{1}})^{2}.  \label{rLE}
\end{equation}
Here, $\lambda_{LE}$ is the reorganization energy for transition from
localized state $A_{1}$ to extended state $B_{2}$. It is interesting to
notice that the activation energy $E_{a}^{LE}$ for LE transition can be
obtained by assuming $\Theta_{\alpha}^{A_{3}}=0$ in $\lambda_{LL}$.
Transition from a localized state to an extended state corresponds to that
of a particle escaping a barrier along the reaction path\cite{kram}.

Formally, the LE transition is similar to LL transition. To obtain the
former, one makes the substitutions: $J_{A_{3}A_{1}}\rightarrow
J_{B_{2}A_{1}}^{\prime}$, $\Theta_{\alpha}^{A_{3}}-\Theta_{\alpha}^{A_{1}}%
\rightarrow \Theta_{\alpha}^{A_{1}}$ and $[(E^{A_{1}}+\mathcal{E}%
_{b}^{A_{1}})-(E^{A_{3}}+\mathcal{E}_{b}^{A_{3}})]\rightarrow\lbrack
E_{B_{2}}-(E_{A_{1}}^{0}+\mathcal{E}_{b}^{A_{1}})]$. However the physical
meaning of the two are different.
~From (\ref{de}) and (\ref{rLE}), we know that $\lambda_{LE}$ is the same
order of magnitude as $\lambda_{LL}$. $\Delta G_{LE}^{0}$ is order of
mobility edge, which is much larger than $\Delta G_{LL}^{0}$ and $%
J^{\prime}_{B_{2}A_{1}}>>J^{\prime}_{A_{3}A_{1}}$. The spatial displacement
of the electron in a LE transition is about the linear size of the localized
state.
~From (\ref{LEmar}) and (\ref{Mar}), $W_{T}^{LE}$ becomes comparable to $%
W_{T}^{LL}$ only when temperature is higher than $%
k_{B}^{-1}(E_{a}^{LE}-E_{a}^{LL})[2\ln(J_{LE}/J_{LL})]^{-1}$. The mobility
edge of a-Si is about $0.1-0.2$eV\cite{big,yue}, so that the LL transition
is dominant in intrinsic a-Si below 580K. However if \textit{higher}
localized states close to the mobility edge are occupied due to doping,
there exist some extended states which satisfy $\Delta G_{LE}^{0}\sim\Delta
G_{LL}^{0}$. For these LE transitions, $E_{a}^{LE}$ is comparable to $%
E_{a}^{LL}$. The LE transition probability is about 10 times larger than
that of LL transition. For these higher localized states, using parameters
given for LL$\ $transition in a-Si, $W_{T}^{LE}\sim10^{13}$sec$^{-1}$.

According to approximation (i): $Y_{BA}<<1$ (Appendix A), (\ref{ple4}) is
only suitable for localized tail states which are far from the mobility
edge. (\ref{ple4}) complements Kikuchi's idea of `phonon induced
delocalization'\cite{kik}: transitions from less localized states close to
mobility edge to extended states. For less localized states, the coupling
with atomic vibrations is weaker\cite{dra,tafn},
the reorganization energy $\lambda_{LE}$ is small. The transition from a
less localized state (close to mobility edge) to an extended state is thus
driven by single-phonon emission or absorption\cite{kik}, similar to the MA
theory\cite{mil}. Consider a localized state and an extended state, both
close to the mobility edge. Then $\Delta G_{LE}^{0}$ is small, $W_{LE}$ can
be large. The inelastic process makes the concept of localization
meaningless for the states close to the mobility edge\cite{thou,im}.
\begin{figure}[tbp]
\resizebox{0.5\textwidth}{!}{  \includegraphics{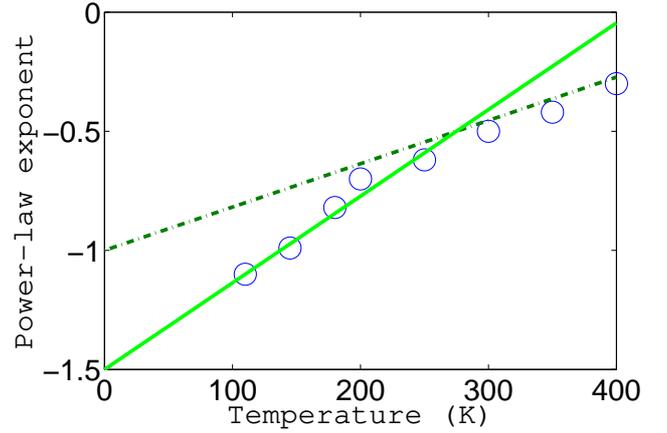}
} 
\caption{photocurrent time decay index as function of temperature: circles
are data for a-Si:H\protect\cite{kris} and a-As$_{2}$Se$_{3}$\protect\cite%
{ore}, dashed line is from the transition probability of MA theory, solid
line is from (\protect\ref{Mar},\protect\ref{LEmar}) in $G<<\protect\lambda$
limit. At higher temperature, $\protect\lambda_{LE}\sim\Delta G_{LE}$,
and the present result reduces to MA theory.}
\label{transient}
\end{figure}


When a gap-energy pulse is applied to amorphous semiconductors, the
transient photocurrent decays with a power-law: $t^{r(T)}$. The exponent is $%
r(T)=-1+k_{B}T/E_{U}$ according to a phenomenological MA type transition
probability, $E_{U}$ is the Urbach energy of the band tail\cite{sch,ore,mon}%
. (\ref{LEmar}) leads to\cite{Hol2} $r(T)=-3/2+2k_{B}T/E_{U}$ if we follow
the reasoning in \cite{ore,mon,bot}. Figure \ref{transient} depicts the
decay index as a function of temperature.
At lower temperature the experimental data deviates from the prediction of
the MA theory.
At higher temperature and for states close to mobility edge, $\lambda
\sim\Delta G$ and $E_{a}\sim\Delta G$, the present theory reduces to MA
theory\cite{mon,bot}.


\section{EL transition and EE transition}

\label{ELEE} If an electron is initially in an extended state $|B_{1}\rangle
$, the amplitude for a transition can be computed in perturbation theory.
The probability of the transition from extended state $|B_{1}\rangle $ to
localized state $|A_{2}\rangle $ is:%
\[
W_{T}(B_{1}\rightarrow A_{2})=\frac{1}{\hbar ^{2}}\exp \{-\frac{\beta }{2}%
(E_{A_{2}}^{0}+\mathcal{E}_{A_{2}}^{b}-E_{B_{1}})\}
\]%
\begin{equation}
\exp \{-\frac{1}{2}\sum_{\alpha }(\theta _{\alpha }^{A_{2}})^{2}\coth \frac{%
\beta \hbar \omega _{\alpha }}{2}\}(I_{1}+I_{2}),  \label{el6}
\end{equation}%
where:%
\[
I_{1}=\int_{-t}^{t}d\tau \exp \{\frac{i\tau }{\hbar }(E_{A_{2}}^{0}+\mathcal{%
E}_{A_{2}}^{b}-E_{B_{1}})\}
\]%
\[
\exp \{\frac{1}{2}\sum_{\alpha }(\theta _{\alpha }^{A_{2}})^{2}csch\frac{%
\beta \hbar \omega _{\alpha }}{2}\cos \omega _{\alpha }\tau \}\times
\]

\[
\lbrack\sum_{\alpha^{\prime}}(\frac{1}{2}(K_{A_{2}B_{1}}^{\prime\alpha
^{\prime}})^{2}+\frac{1}{4}(K_{A_{2}B_{1}}^{\prime\alpha^{\prime}}\theta_{%
\alpha^{\prime}}^{A_{2}})^{2}\coth\frac{\beta\hbar\omega _{\alpha^{\prime}}}{%
2})
\]

\[
\frac{\hbar}{M_{\alpha^{\prime}}\omega_{\alpha^{\prime }}}csch\frac{%
\beta\hbar\omega_{\alpha^{\prime}}}{2}\cos\omega_{\alpha^{\prime }}\tau
\]

\begin{equation}
-\frac{1}{4}\sum_{\alpha ^{\prime }}\frac{\hbar }{M_{\alpha ^{\prime
}}\omega _{\alpha ^{\prime }}}(K_{A_{2}B_{1}}^{\prime \alpha ^{\prime
}}\theta _{\alpha ^{\prime }}^{A_{2}})^{2}csch^{2}\frac{\beta \hbar \omega
_{\alpha ^{\prime }}}{2}\cos ^{2}\omega _{\alpha ^{\prime }}\tau ],
\label{i1}
\end{equation}%
and:%
\[
I_{2}=[\frac{1}{4}\sum_{\alpha ^{\prime }}(K_{A_{2}B_{1}}^{\prime \alpha
}\Theta _{\alpha ^{\prime }}^{A_{2}})^{2}]\int_{-t}^{t}d\tau \exp \{\frac{%
i\tau }{\hbar }(E_{A_{2}}^{0}+\mathcal{E}_{A_{2}}^{b}-E_{B_{1}})\}\times
\]%
\begin{equation}
\lbrack \exp \{\frac{1}{2}\sum_{\alpha }(\theta _{\alpha }^{A_{2}})^{2}csch%
\frac{\beta \hbar \omega _{\alpha }}{2}\cos \omega _{\alpha }\tau \}-1].
\label{i2}
\end{equation}

For very high temperature $k_{B}T\geq 2.5\hbar \overline{\omega }$, the EL
transition probability is%
\begin{equation}
W_{EL}=\nu _{EL}e^{-E_{a}^{EL}/k_{B}T},  \label{elp}
\end{equation}%
with%
\[
\nu _{EL}=\frac{1}{\hbar }(\frac{\pi }{k_{B}T\lambda _{EL}})^{1/2}\{\frac{1}{%
4}\sum_{\alpha ^{\prime }}(K_{A_{2}B_{1}}^{\prime \alpha ^{\prime }}\Theta
_{\alpha ^{\prime }}^{A_{2}})^{2}
\]%
\[
+\sum_{\alpha ^{\prime }}[\frac{(K_{A_{2}B_{1}}^{\prime \alpha ^{\prime
}})^{2}\hbar }{2M_{\alpha ^{\prime }}\omega _{\alpha ^{\prime }}}csch\frac{%
\beta \hbar \omega _{\alpha ^{\prime }}}{2}+\frac{(K_{A_{2}B_{1}}^{\prime
\alpha }\Theta _{\alpha ^{\prime }}^{A_{2}})^{2}}{8}sech^{2}\frac{\beta
\hbar \omega _{\alpha ^{\prime }}}{4}]
\]

\[
-[2(\sum_{\alpha}\omega_{\alpha}^{2}(\theta_{\alpha}^{A_{2}})^{2}csch\frac{%
\beta\hbar\omega_{\alpha}}{2})^{-1}
\]

\[
-4\frac{(E_{A_{2}}^{0}+\mathcal{E}_{A_{2}}^{b}-E_{B_{1}})^{2}}{%
(\sum_{\alpha}\hbar\omega_{\alpha }^{2}(\theta_{\alpha}^{A_{2}})^{2}csch%
\frac{\beta\hbar\omega_{\alpha}}{2})^{2}}]\times
\]

\[
\times\sum_{\alpha^{\prime}}[\frac{(K_{A_{2}B_{1}}^{\prime\alpha^{%
\prime}})^{2}\hbar\omega_{\alpha^{\prime}}}{4M_{\alpha^{\prime}}}csch\frac{%
\beta \hbar\omega_{\alpha^{\prime}}}{2}
\]

\[
+\frac{\omega_{\alpha^{\prime}}^{2}(K_{A_{2}B_{1}}^{\prime\alpha^{\prime}}%
\Theta_{\alpha^{\prime}}^{A_{2}})^{2}\coth\frac{\beta\hbar\omega_{\alpha^{%
\prime}}}{2}}{8}csch\frac{\beta \hbar\omega_{\alpha^{\prime}}}{2}
\]

\begin{equation}
-\frac{\omega _{\alpha ^{\prime }}^{2}(K_{A_{2}B_{1}}^{\prime \alpha
^{\prime }}\Theta _{\alpha ^{\prime }}^{A_{2}})^{2}}{4}csch^{2}\frac{\beta
\hbar \omega _{\alpha ^{\prime }}}{2}]\},  \label{nuel}
\end{equation}%
and%
\[
E_{a}^{EL}=\frac{\lambda _{EL}}{4}(1+\frac{\Delta G_{EL}^{0}}{\lambda _{EL}}%
)^{2},~~\lambda _{EL}=\frac{1}{2}\sum_{\alpha }M_{\alpha }\omega _{\alpha
}^{2}(\Theta _{\alpha }^{A_{2}})^{2},
\]%
\begin{equation}
\Delta G_{EL}^{0}=E_{A_{2}}^{0}+\mathcal{E}_{A_{2}}^{b}-E_{B_{1}}<0,
\label{acel}
\end{equation}%
where $\lambda _{EL}$ is order of magnitude of $\lambda _{LL}$. Because $\Delta
G_{EL}^{0}<0$, from the expressions of $E^{EL}_{a}$ and $E^{LL}_{a}$, we know $E_{a}^{EL}$ is smaller than $E_{a}^{LL}$. Since $%
K_{A_{2}B_{1}}^{\prime \alpha ^{\prime }}\Theta _{\alpha ^{\prime }}^{A_{2}}$
is the same order magnitude as $J_{A_{3}A_{1}}$, the EL transition
probability is larger than that of LL transition.
In a-Si, this yields $W_{T}^{EL}\sim 10^{13}-10^{14}$sec$^{-1}$. When a
carrier moves in a localized state, the nearby atoms are shifted due to the
e-ph interaction, so that all modes are affected. Thus an EL transition is a
multi-phonon process.

\begin{figure}[tbp]
\resizebox{0.5\textwidth}{!}{  \includegraphics{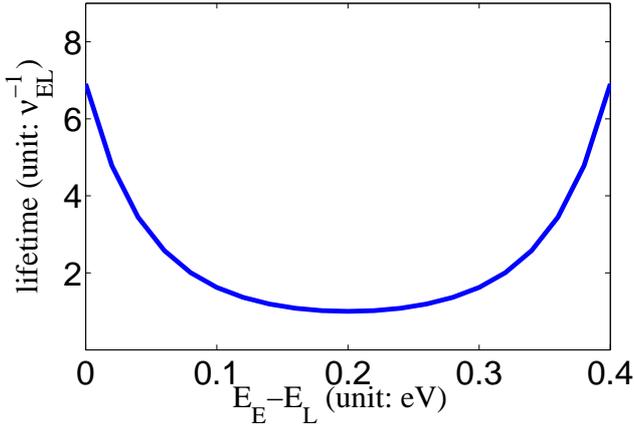}
} 
\caption{The non-radiative transition lifetime of extended state as function
of energy difference $E_{E}-E_{L}$ between initial extended state and final
localized state. $\protect\lambda_{EL}=0.2$ eV is estimated from the data
given in Sec. III. When $E_{E}-E_{L}=\protect\lambda_{EL}$, lifetime is a
minimum. The short lifetime belt exists in range 0.1 eV$<E_{E}-E_{L}<0.3$eV.
The vertical axis is scaled by $\protect\nu_{EL}^{-1}$.}
\label{belt}
\end{figure}

$\Delta G_{EL}^{0}<0$ has a deep consequence. From the expression (\ref{ev})
for $\mathcal{E}_{A_{2}}^{b}$ and the order of magnitude of mobility edge%
\cite{sca}, we know that the energy difference $\Delta G_{EL}^{0}<0$ is
order several tenths eV. For extended states with $|\Delta
G_{EL}^{0}|/\lambda_{EL}<1$, we are in the normal regime: the higher the
energy of an extended state (i.e. $\Delta G_{EL}^{0}/\lambda_{EL}$ more
negative but still $|\Delta G_{EL}^{0}|/\lambda_{EL}<1$), the smaller the
activation energy $E_{a}^{EL}$. The higher extended state has a shorter
lifetime, therefore the time that an electron is able to remain in such an
extended state is less than the time it spends in a lower extended state. An
extended state with shorter lifetime contributes less to the conductivity.
For extended states with energies \textit{well above} the mobility edge
(such that $|\Delta G_{EL}^{0}|/\lambda_{EL}>1$), we are in Marcus inverted
regime (cf. Figure \ref{belt}): the higher the energy of an extended state,
the larger the activation energy. The higher extended states have long
lifetimes and will contribute more to conductivity. In the middle of the two
regimes, $\Delta G_{EL}^{0}/\lambda_{EL}\approx-1$. For these extended
states, \textit{no activation energy is required for the transition to
localized states}. Such extended states will quickly decay to the localized states. In
experiments, there is indirect evidence for the existence of this
short-lifetime belt. In a crystal, phonon-assisted non-radiative transitions
are slowed by the energy-momentum conservation law. In c-Si/SiO$_{2}$
quantum well structure, the photoluminescence lifetime is about 1 ms, and is
insensitive to the wavelength\cite{oka}. The photoluminescence lifetime of
a-Si/SiO$_{2}$ structure becomes shorter with a decrease in wavelength: 13ns
at 550nm and 143ns at 750nm\cite{kan}. The trend is consistent with the left
half of Figure \ref{belt}. The observed wavelength indicates that the energy
difference ($>1.66$eV) between the hole and electron is larger than band gap
(1.2eV), so that the excited electrons are in extended states. According to (%
\ref{nuel}) and Figure \ref{belt}, higher extended states are more quickly
depleted by the non-radiative transitions than the lower ones, so that a
photoluminescence signal with higher frequency has a shorter lifetime. We
need to be careful on two points: (i) the observed recombination time is
order of ns, it is the EE transitions that limits EL transition to a large
extent; (ii) for a quantum well, the number of atoms is small, so that the
reorganization energy is smaller than the bulk. The static displacements may
be able to adjust at the experiment temperature 2-10K. To really prove the
existence of the short-lifetime belt, one needs to excite electrons into and
above the belt with two narrow pulses: if the higher energy luminescence
lasts longer than the lower energy one, the existence of a short-lifetime
belt is demonstrated.

In the conduction band, the energy of any localized tail state is lower than
that of any extended state, so that a zero-phonon process is impossible.
Because the $\Delta G_{EL}^{0}<0,$ factor $\exp [|\Delta
G_{EL}^{0}|/(2k_{B}T)]$ increases with decreasing temperature. On the other
hand, other factors in (\ref{loweL}) decrease with decreasing temperature.
Thus there exists an optimal temperature T$_{\ast }$, at which the
transition probability is maximum. If one measures the variation of
luminescence changing with temperature in low temperature region, at T$%
_{\ast }$ the lifetime of the photoluminescence will be shortest.

A transition from one extended state to another extended state is a
single-phonon absorption or emission process driven by e-ph interaction. The
transition probability from extended state $|B_{1}\rangle$ to extended state
$|B_{2}\rangle$ is:

\[
W(B_{1}\rightarrow B_{2})=\frac{2\pi }{\hbar }\sum_{\alpha ^{\prime
}}(K_{B_{2}B_{1}}^{\alpha ^{\prime }})^{2}\frac{\hbar }{M_{\alpha ^{\prime
}}\omega _{\alpha ^{\prime }}}
\]%
\begin{equation}
\lbrack \frac{\overline{N_{\alpha ^{\prime }}^{\prime }}}{2}\delta
(E_{B_{2}}-E_{B_{1}}-\hbar \omega _{\alpha ^{\prime }})+\frac{\overline{%
N_{\alpha ^{\prime }}^{\prime }}+1}{2}\delta (E_{B_{2}}-E_{B_{1}}+\hbar
\omega _{\alpha ^{\prime }})],  \label{avie}
\end{equation}%
where $\overline{N_{\alpha ^{\prime }}^{\prime }}=(e^{\beta \hbar \omega
_{\alpha ^{\prime }}}-1)^{-1}$ is the average phonon number in the $\alpha
^{\prime }$th mode. In a crystal, (\ref{avie}) arises from inelastic
scattering with phonons.

\subsection{Four transitions and conduction mechanisms}

\label{4t}

\begin{center}
\begin{table*}[ht]
\caption{Some features of 4 types of transitions in a-Si.}
\label{tab:4tran}{\small \hfill{}
\begin{tabular}{lllllll}
\hline\hline
transitions & origin & phonons needed & activated & role in conduction &
probability $W_{T}$(sec$^{-1}$) &  \\ \hline
L$\rightarrow$L & $J$ (\ref{tra1}) & multi & \ \ \ yes & direct & $10^{12}$
&  \\ \hline
L$\rightarrow$E & $J^{^{\prime}}$ (\ref{tra2}) & multi & \ \ \ yes &
direct+indirect & $10^{10}-10^{13}$ &  \\ \hline
E$\rightarrow$L & $K^{^{\prime}}$ (\ref{ept}) & multi & \ \ \ yes &
direct+indirect & $10^{13}-10^{14}$ &  \\ \hline
E$\rightarrow$E & $K$ (\ref{ep0}) & single & \ \ \ no & reduce & $10^{13}$ &
\\ \hline\hline
\end{tabular}%
} \hfill{}
\end{table*}
\end{center}

The characteristics of the four types of transitions are summarized in Table \ref{tab:4tran}%
. The last column gives the order of magnitude of the transition probability
estimated from the parameters of a-Si at T=300K. In a-Si:H, the role of hydrogen atoms is to passivate dangling bond, the estimated rates are roughly applicable to a-Si:H. The rate of LL transitions is between two
nearest neighbors. For an intrinsic or lightly n-doped semiconductor at
moderate temperature (for a-Si $T<580$K, the energy of mobility edge), only
the lower part of conduction tail is occupied. Then $\Delta G_{LE}$ is
large, and the LE transition probability is about two orders of magnitude
smaller than that of the LL\ transition. For an intrinsic semiconductor at
higher temperature or a doped material, $\Delta G_{LE}^{0}$ becomes
comparable to $\Delta G_{LL}^{0}$, LE transition probability is about ten
times larger than that of LL transition. The first three transitions
increase the mobility of an electron, whereas EE transition decreases
mobility of an electron.

Although $W_{EL}>W_{LE}$, the transient decay of photocurrent is still
observable\cite{ore}. The reason is that for extended states below the
short-lifetime belt, $W_{EL}\sim W_{EE}$, so that an electron in an extended
state can be scattered into another extended state and continue to
contribute to conductivity before becoming trapped in some localized state.

If a localized tail state is close to the bottom of the conduction band, for
another well localized state and an extended state close to mobility edge, $%
\Delta G_{LE}\sim \Delta G_{LL}+D$, and $W_{LE}$ is one or two orders of
magnitude smaller than $W_{LL}$. If a localized state is close to mobility
edge, $\Delta G_{LE}\sim \Delta G_{LL},$ because $J^{\prime }$ is several
times larger than $J,$ $W_{LE}$ could be one order of magnitude larger than $%
W_{LL}$. The probability of EL transition is then one or two orders of
magnitude larger than that of LL transition. The reason is that $\Delta
G_{EL}<0$, $E_{a}^{EL}$ is smaller than $E_{a}^{LL}$ while $K^{\prime
}\Theta $ is the same order of magnitude as $J$. The probability of EE
transition $W_{EE}$ is about 10$^{3}$ times larger than $W_{LL}$ (cf. Table %
\ref{tab:4tran}). The EE transition deflects the drift motion which is along
the direction of electric field and reduces conductivity. This is in contrast
with the LL, LE and EL transitions. The relative contribution to
conductivity of four transitions also depends on the number of carriers in
extended states and 
in localized states, which are determined by the extent of doping and
temperature. At low temperature ( $k_{B}T<\hbar \overline{\omega }/10$), the
non-diagonal transition is still multi-phonon activated process whatever it
is LL, LE or EL transition. The activation energy is just half of the energy
difference between final state and initial state (cf. (\ref{de}), (\ref{ELd}%
) and (\ref{acel})).

\subsection{Long time and higher order processes}

\label{ho} The perturbation treatments of the four fundamental processes are
only suitable for short times, in which the probability amplitude of the
final state is small. Starting from a localized state we only have $%
L\rightarrow E$ process and $L\rightarrow L$ process. Starting from an
extended state, we\ only have $E\rightarrow L$ process and $E\rightarrow E$
process. For long times, higher order processes appear. For example $%
L\rightarrow E\rightarrow L\rightarrow L\rightarrow E\rightarrow
L\rightarrow E\rightarrow E\rightarrow L$ etc. Those processes are important
in amorphous solids. In a macroscopic sample, there are many occupied
localized and extended states. If we are concerned with the collective
behavior of all carriers rather than an individual carrier in a long time
period, the picture of the four transitions works well statistically.


All four transitions are important to dc conductivity and transient
photocurrent. In previous phenomenological models, the role of LE transition
was taken into account by parameterizing the MA probability\cite{ore,mon}.
The details of LE transition and the polarization of network by the
localized carriers were ignored. The present work has attempted to treat the
four transitions in a unified way. Our approach enhanced previous theories
in two aspects: (1) the role of polarization is properly taken into account;
and (2) we found the important role played by the EL transition and
associated EE transition in dc conductivity and in the non-radiative decay
of extended states.

\section{Summary}

\label{sum} For amorphous solids, we established the evolution equations for
localized tail states and extended states in the presence of lattice
vibrations.
For short times, perturbation theory can be used to solve Eqs.(\ref{s11})
and (\ref{s22}). The transition probabilities of LL, LE, EL and EE
transitions are obtained. The relative rates for different processes and the
corresponding control parameters are estimated.

The new results found in this work are summarized in the following. At high temperature, any transition involving well-localized state(s) is a multi-phonon process, the transition rate takes form (\ref{ea}). At low temperature, variable range hopping appears as the most probable LL transitions.

The field-dependence of the conductivity estimated from LL transition is closer
to experiments than previous theories.
The predicted Meyer-Neldel temperature and the linear relation between the
upper temperature limit of VRH and the frequency $\bar{\nu}$ of first peak
of phonon spectrum are consistent with experiments in quite different
materials.

We suggested that there exists a short lifetime belt of
extended states inside conduction band or valence band. These states favor
non-radiative transitions by emitting several phonons.

From (\ref{sl}) one can see that single-phonon LL transition appears when states become less localized.
 In intrinsic or lightly doped amorphous semiconductors,\ the well localized
states are the low lying excited states and are important for transport.
Carriers in these well localized tail states polarize the network: any
process involving occupation changes of well localized tail states must change the
occupation numbers in many vibrational modes and are multi-phonon process.
Moving towards to the mobility edge, the localization length of a state
becomes larger and larger. When the static displacements caused by
the carrier in a less localized state are comparable to the vibrational
displacements, one can no longer neglected the vibrational displacements in (%
\ref{sl}). The usual electron-phonon interaction also plays a role in causing
transition from a less localized state. In this work, we did not discuss
this complicated situation. %
Formally when reorganization energy $\lambda _{LL}$ between two localized
states is small and comparable to the typical energy difference $\Delta
G_{LL}$, the present multi-phonon LL transition probability reduces to
single-photon MA theory.

The multi-phonon LE transition discussed in this
work is for well localized states, it is a supplement to the theory of
phonon-induced delocalization which is concerned with the less localized states close
to the mobility edge. As discussed in Appendix A, when the atomic static displacements caused by a carrier in a less localized state is comparable to the vibrational amplitude, LE transition could be caused either by a phonon in resonance with the initial and final states or by the transfer integral $J^{'}_{B_{2}A_{1}}$.

\begin{acknowledgement}
We thank the Army Research Office for support under MURI W91NF-06-2-0026, and
the National Science Foundation for support under grants DMR 0903225.
\end{acknowledgement}

\appendix

\section{Approximations used to derive evolution equations}

\label{der} Usually in the zero order approximation of crystals (specially
in the theory of metals), the full potential energy \\$\sum_{\mathbf{n}}U_{\mathbf{%
n}}(r-\mathbf{R}_{\mathbf{n}}-\mathbf{u}_{v}^{\mathbf{n}})$ is replaced by $%
V_{c}=\sum_{\mathbf{n}}U_{\mathbf{n}}(r-\mathbf{R}_{\mathbf{n}})$, where $%
\mathbf{R}_{\mathbf{n}}$ and $\mathbf{u}_{v}^{\mathbf{n}}$ are the
equilibrium position and vibrational displacemet of the $n$th atom. One then
diagonalizes $h_{c}=-\hbar ^{2}\nabla ^{2}/2m+V_{c}$, such that all eigenstates are
orthogonal. Electron-phonon interaction $\sum_{j=1}^{3\mathcal{N}}x_{j}\frac{%
\partial U}{\partial X_{j}}$\ slightly modifies the eigenstates and
eigenvalues of $h_{c}$ or causes scattering between eigenstates. This
procedure works if e-ph interaction does not fundamentally change the nature
of Bloch states. As long as static disorder is not strong enough to cause
localization, Bloch states\ are still proper zero order states.
However one should be careful when dealing with the effect of static
disorder $\{\mathbf{u}_{s}^{\mathbf{n}}\}$. It must be treated as
perturbation along with e-ph interaction. If we put static disorder in
potential energy and diagonalize $h_{s}=-\hbar ^{2}\nabla ^{2}/2m+\sum_{%
\mathbf{n}}U_{\mathbf{n}}(r-\mathbf{R}_{\mathbf{n}}-\mathbf{u}_{s}^{\mathbf{n%
}})$, the scattering effect of static disorder disappear in the
disorder-dressed eigenstates. The physical properties caused by static
disorder e.g. resistivity is not easy to display in an intuitive kinetic
consideration based on Boltzmann-like equation: eigenstates of $h_{s}$ are
not affected by static disorder $\{\mathbf{u}_{s}^{\mathbf{n}}\}$. By contrast,
computing transport coefficients with eigen states of $h_{s}$ is not a
problem in Kubo formula or its improvment\cite{kubo}.

We face a dilemma in amorphous semiconductors. On one hand the static
disorder $\{\mathbf{u}_{s}^{\mathbf{n}}\}$ is so strong that some band tail
states are localized, static disorder must be taken into account at zero
order i.e. diagonalize $h_{s}$; on the other hand from kinetic point-of-view
the carriers in extended states are scattered by the static disorder which
should be displayed explicitly $\sum_{\mathbf{n}}\mathbf{u}_{s}^{\mathbf{n}%
}\cdot \partial U_{\mathbf{n}}(r-\mathbf{R}_{\mathbf{n}})/\partial \mathbf{R}%
_{\mathbf{n}}$ rather than included in the exact eigenstates of $h_{s}$.


The very different strengths of the e-ph interaction in localized states and in
extended states also requires different partitions of the potential energy $%
\sum_{\mathbf{n}}U_{\mathbf{n}}(r-\mathcal{R}_{\mathbf{n}}-\mathbf{u}_{v}^{%
\mathbf{n}})$, where $\mathcal{R}_{\mathbf{n}}=\mathbf{R}_{\mathbf{n}}+%
\mathbf{u}_{s}^{\mathbf{n}}$ is the static position of the $n^{th}$ atom in an
amorphous solid. Molecular dynamics (MD) simulations\cite{dra,tafn} show
that the eigenvalues of localized states are strongly modified (about
several tenth eV) by e-ph interaction, while the eigenvalues of extended
states do not fluctuate much. It seems reasonable that for localized states
we should include e-ph interaction at zero order, and put it in the zero-order single-particle potential
energy (just like small polaron theory\cite{Hol1,Hol2}), while for extended
states e-ph interaction acts like a perturbation (similar to the inelastic
scattering of electrons caused by e-ph interaction in metals).

The traditional partition of full potential energy is%
\begin{equation}
\sum_{\mathbf{n}}U(r-\mathcal{R}_{\mathbf{n}},\mathbf{u}^{\mathbf{n}})=\sum_{%
\mathbf{n}=1}^{\mathcal{N}}U(\mathbf{r}-\mathcal{R}_{\mathbf{n}%
})+\sum_{j=1}^{3\mathcal{N}}x_{j}\frac{\partial U}{\partial X_{j}}.
\label{old}
\end{equation}%
In this \textit{ansatz}, static disorder is included at zero order. Localized states
and extended states are eigen states of $h_{s}$. The second term of (\ref%
{old}), the e-ph interaction, is the unique residual perturbation to eigenstates of $h_{s}$. It causes transitions among the eigen states of $h_{s}$
i.e. LL, LE, EL and EE transitions, to lowest order transition is driven by
single-phonon absorption or emission. Since all attractions due to static
atoms are included in $h_{s}$, two types of transfer integrals $%
J_{A_{2}A_{1}}$ (from a localized state to another localized state) and $%
J_{B_{2}A_{1}}^{\prime }$ (from a localized state to an extended state) do
not exist.
One can still use Kubo formula or subsequent development\cite{kubo} to compute
transport coefficients.

However partition (\ref{old}) obscures the construction of a \\Boltzmann-like
picture for electronic conduction where various agitation and obstacle
mechanisms are explicitly exposed. The elastic scattering caused by static
disorder is hidden in the eigen states of $h_{s}$. To explicitly illustrate
the elastic scattering produced by\ static disorder, one has to further
resolve the first term of (\ref{old}) into%
\[
\sum_{\mathbf{n}=1}^{\mathcal{N}}U(\mathbf{r}-\mathcal{R}_{\mathbf{n}%
})=\sum_{\mathbf{n}=1}^{\mathcal{N}}U(\mathbf{r}-\mathbf{R}_{\mathbf{n}})
\]%
\begin{equation}
+\sum_{\mathbf{n}}\mathbf{u}_{s}^{\mathbf{n}}\cdot \partial U_{\mathbf{n}}(r-%
\mathbf{R}_{\mathbf{n}})/\partial \mathbf{R}_{\mathbf{n}}.  \label{stadis}
\end{equation}%
(\ref{old}) is also inconvenient for localized states. The e-ph interaction
for a carrier in a localized state is much stronger than in an extended state%
\cite{dra,tafn}. It is reflected in two aspects: (i) a localized carrier
polarizes network and produces static displacements for the atoms in which
the localized state spread; (ii) the wave functions and corresponding
eigenvalues of localized states are obviously changed (the change in
eigenvalues can be clearly seen in MD trajectory\cite{dra,tafn}). To
describe these two effects, in perturbation theory one has to calculate e-ph
interaction to infinite order.

Taking different partitions for localized states and extended states is a
practical \textit{ansatz}.%
For a localized state, we separate%
\[
\sum_{\mathbf{n}}U(r-\mathcal{R}_{\mathbf{n}},\mathbf{u}_{v}^{\mathbf{n}%
})=\sum_{\mathbf{n}\in D_{A_{1}}}U(r-\mathcal{R}_{\mathbf{n}},\mathbf{u}%
_{v}^{\mathbf{n}})
\]%
\begin{equation}
+\sum_{\mathbf{p}\notin D_{A_{1}}}U(r-\mathcal{R}_{\mathbf{p}},\mathbf{u}%
_{v}^{\mathbf{p}}),  \label{sl}
\end{equation}%
where $D_{A_{1}}$ is the distorted region where localized tail state $\phi
_{A_{1}}$ spreads. For a nucleus outside $D_{A_{1}}$, its effect on
localized state $A_{1}$ dies away with the distance between the nucleus and $%
D_{A_{1}}$. The second term leads to two transfer integrals $J_{A_{2}A_{1}}$
(induces LL transition) and $J_{B_{2}A_{1}}^{\prime }$ (induces LE
transition) in the evolution equations of localized states.

Since for a well-localized state $\phi _{A_{1}}$, the wave function is only
spread on the atoms in a limited spatial region $D_{A_{1}}$, $\langle \phi
_{A_{1}}|\cdot |\phi _{A_{1}}\rangle =0$, where $\cdot $ stands for the
second term in the RHS of (\ref{sl}). In calculating $J_{A_{2}A_{1}}=\langle
\phi _{A_{2}}|\cdot |\phi _{A_{1}}\rangle $ and $J_{B_{2}A_{1}}^{\prime
}=\langle \xi _{B_{2}}|\cdot |\phi _{A_{1}}\rangle $, it is legitimate to
neglect $\mathbf{u}_{v}^{\mathbf{p}}$, $\cdot \thickapprox \sum_{\mathbf{p}%
\notin D_{A_{1}}}U(r-\mathcal{R}_{\mathbf{p}})$. The change in potential
energy induced by the atomic vibrational displacements is fully included in
the first term in RHS of (\ref{sl}). Because the wave function $\phi _{A_{1}}
$ of localized state $A_{1}$ is confined in $D_{A_{1}}$, one can view $\phi
_{A_{1}}$ as the eigenfunction of $h_{A_{1}}^{0}=-\hbar ^{2}\nabla
^{2}/2m+\sum_{\mathbf{n}\in D_{A_{1}}}U(r-\mathcal{R}_{\mathbf{n}},\mathbf{u}%
_{v}^{\mathbf{n}})$ with eigenvalue $E_{A_{1}}(\{\mathbf{u}_{v}^{\mathbf{n}},%
\mathbf{n}\in D_{A_{1}}\})$. The rest of atoms outside $D_{A_{1}}$ act as
boundary of $\phi _{A_{1}}$. A carrier in localized state $\phi _{A_{1}}$
propagates in region $D_{A_{1}}$ and is reflected back at the boundary of $%
D_{A_{1}}$.
Since static disorder is fully contained in $h_{A_{1}}^{0}$, in the present
\textit{ansatz}, localized carriers are free from elastic scattering of static
disorder.

For a less localized state $\phi _{A_{1}}$, its wave function spreads over a
wider spatial region $D_{A_{1}}$. With shift to the mobility edge, $\sum_{%
\mathbf{n\in }D_{A_{1}}}\mathbf{u}_{v}^{\mathbf{n}}\cdot \nabla U(r-\mathcal{%
R}_{\mathbf{n}})$ become smaller and smaller, eventually comparable to $%
\sum_{\mathbf{p}\notin D_{A_{1}}}U(r-\mathcal{R}_{\mathbf{p}})$ and $\sum_{%
\mathbf{p}\notin D_{A_{1}}}\mathbf{u}_{v}^{\mathbf{p}}\cdot \nabla U(r-%
\mathcal{R}_{\mathbf{p}},\mathbf{u}_{v}^{\mathbf{p}})$. For carriers on
these less localized states, their polarization of the network is weak, and the
atomic static displacements are comparable to the vibrational amplitudes.
Entering or leaving a less localized state does not require configuration
reorganization, and the reorganization energy becomes same order of magnitude as
vibrational energy. For such a situation, only a phonon in resonance with
the initial and final states contributes to the transition. The multi-phonon
processes gradually become the single-phonon processes, although the driving
force is still the transfer integral induced by $\sum_{\mathbf{p}\notin
D_{A_{1}}}U(r-\mathcal{R}_{\mathbf{p}})$. One obtains the MA theory.

For a less localized state, if we treat $\sum_{%
\mathbf{n\in }D_{A_{1}}}\mathbf{u}_{v}^{\mathbf{n}}\cdot \nabla U(r-\mathcal{%
R}_{\mathbf{n}})$, $%
\sum_{\mathbf{p}\notin D_{A_{1}}}U(r-\mathcal{R}_{\mathbf{p}})$ and $\sum_{%
\mathbf{p}\notin D_{A_{1}}}\mathbf{u}_{v}^{\mathbf{p}}\cdot \nabla U(r-%
\mathcal{R}_{\mathbf{p}},\mathbf{u}_{v}^{\mathbf{p}})$ in the same foot as perturbation, the phonon-induced delocalization naturally appears and is accompany with EL transitions induced by transfer integral $J^{'}_{B_{2}A_{1}}$.

For extended states, to construct a kinetic description, (\ref{old}) is
suitable partition of full potential energy. 
The second term of (\ref{old}), the e-ph interaction, causes EL and EE
transitions in the evolution equation of extended state. The elastic
scattering of the carriers in extended states induced by static disorder can
be taken into account by two methods: (i) Using the eigenvalues and eigenfunctions of $h_{s}$ in Kubo formula; (2) if we wish to deal static disorder
more explicitly, we can apply coherent potential approximation to (\ref%
{stadis}). In this work we do not discuss these issues and only concentrate
on the transitions involving atomic vibrations.

In deriving (\ref{s11},\ref{s22}), we neglected the dependence of extended
state $\xi_{B_{1}}$ on the vibrational displacements, then $%
\nabla_{j}\xi_{B_{1}}=0$. Since $v_{n}/v_{e}\sim10^{-3}$ (where $v_{e}$ and $%
v_{n}$ are typical velocities of the electron and nucleus), $m/M\sim10^{-4}$
($m$ and $M$ are mass of electron and of a typical nucleus) and $%
x/d\sim10^{-2}-10^{-1}$ ($x$ and $d$ are typical vibrational displacement of
atom and bond length), one can show $\sum_{j}(\hbar^{2}/M_{j})(%
\nabla_{j}a_{A_{1}})(\nabla_{j}\phi_{A_{1}})$ or $\sum_{j}(\hbar^{2}/2M_{j})%
\nabla_{j}^{2}\phi_{A_{1}}<<\sum_{\mathbf{p}\notin D_{A_{1}}}U(r-\mathcal{R}%
_{\mathbf{p}},\mathbf{u}^{\mathbf{p}})\phi_{A_{1}}$ or $\sum_{j}x_{j}(%
\partial U/\partial X_{j})\xi_{B_{1}}$. Therefore terms including $%
\nabla_{j}\phi_{A_{1}}$ or \ $\nabla_{j}^{2}\phi_{A_{1}}$ can be neglected.

To further simplify the evolution equations, we need two connected technical
assumptions: (i) the overlap integral $Y_{B_{2}A_{1}}$ between extended
state $\xi_{B_{2}}$ and localized tail state $\phi_{A_{1}}$ satisfies $%
Y_{B_{2}A_{1}}\sim N_{A_{1}}/\mathcal{N}<<1$ and (ii) overlap integral $%
S_{A_{2}A_{1}}$ between two localized tail states satisfies $%
S_{A_{2}A_{1}}<<1$. Assumption (i) means that we do not consider the
localized tail states very close to the mobility edge and consider only the
most localized tail states. Condition (ii) is satisfied for two localized
states which do not overlap. It means we exclude the indirect contribution
to conductivity from the transitions between two localized tail states with
overlapping spatial regions.

For two localized states $A_{2}$ and $A_{1}$, $S_{A_{2}A_{1}}<<1$ if $%
D_{A_{1}}$ and $D_{A_{2}}$ do not overlap. The terms multiplied by $%
S_{A_{2}A_{1}}$ can be neglected for localized tail states which their
spatial regions do not overlap. What is more, the transfer integral is
important only when the atoms $\mathbf{p}\notin D_{A_{1}}$ fall into $%
D_{A_{2}}$ or $A_{1}=A_{2}$,%
\[
\sum_{A_{1}}a_{A_{1}}\int d^{3}r\phi _{A_{2}}^{\ast }\sum_{\mathbf{p}\notin
D_{A_{1}}}U(r-\mathcal{R}_{\mathbf{p}},\mathbf{u}^{\mathbf{p}})\phi _{A_{1}}
\]

\begin{equation}
\approx \sum_{A_{1}}a_{A_{1}}J_{A_{2}A_{1}}+W_{A_{2}}a_{A_{2}}  \label{tra}
\end{equation}%
where $W_{A_{2}}=\int d^{3}r|\phi _{A_{2}}|^{2}\sum_{\mathbf{p}\notin
D_{A_{2}}}U(r-\mathcal{R}_{\mathbf{p}},\mathbf{u}^{\mathbf{p}})$ only
affects the self energy of a localized state through $a_{A_{2}}$. Comparing
with $E_{A_{1}}$ and with $h_{v}$, $W_{A_{2}}$ can be neglected\cite{Hol1}.

\section{Scales of the coupling parameters of four transitions}

\label{para} For these most localized tail states, the wave functions take
the form of $\phi_{A_{1}}\sim e^{-|\mathbf{r}-\mathcal{R}_{A_{1}}|/\xi_{1}}$%
. $J_{A_{2}A_{1}}$ is estimated to be $-(N_{A_{2}}Z^{\ast}e^{2}/4\pi\epsilon
_{0}\varepsilon_{s}\xi)(1+R_{12}/\xi)e^{-R_{12}/\xi}$, where average
localization length $\xi$ is defined by $2\xi^{-1}=\xi_{1}^{-1}+\xi_{2}^{-1}$%
. $R_{12}$ is the average distance between two localized states, $%
\varepsilon _{s}$ is the static dielectric constant, $Z^{\ast}$ is the
effective nuclear charge of atom, $N_{A_{2}}$ is the number of atoms inside
region $D_{A_{2}}$. Later we neglect the dependence of $J_{A_{2}A_{1}}$ on
the vibrational displacements $\{x\}$ of the atoms and consider $%
J_{A_{2}A_{1}}$ as a function of the distance $R_{12}$ between two localized
states, localization length $\xi_{1}$ of state $\phi_{A_{1}}$ and
localization length $\xi_{2}$ of state $\phi_{A_{2}}$. If a localized state
is not very close to the mobility edge, its localization length $\xi$ is
small. The overlap between it and an extended state $Y_{B_{2}A_{1}}\sim
N_{A_{1}}/\mathcal{N}$ may be neglected.

If we approximate extended states as plane waves $\xi _{B_{1}}\sim
e^{ik_{B_{1}}r}$, then $K_{A_{2}B_{1}}^{\prime }\sim (Z^{\ast }e^{2}u/4\pi
\epsilon _{0}\varepsilon _{s}\xi _{A_{2}}^{2})(1-ik_{B_{1}}\xi
_{A_{2}})^{-1} $. So that $K_{A_{2}B_{1}}^{\prime }u/J_{A_{2}A_{1}}\sim
e^{R_{12}/\xi }u/\xi $, where $u\sim \\ \sqrt{k_{B}T/M\omega ^{2}}$ or $\sqrt{%
\hbar /M\omega } $ is typical amplitude of vibration at high or low
temperature. The distance between two nearest localized states is $\sim$
several \AA\ in a-Si, and $K_{A_{2}B_{1}}^{\prime }u$ is several times smaller
than $J_{A_{2}A_{1}}$. If we again approximate extended state as plane wave $%
\xi _{B_{2}}^{\ast }\sim e^{-ik_{B_{2}}r}$, $J_{B_{2}A_{1}}^{\prime }\sim
(Z^{\ast }e^{2}/4\pi \epsilon _{0}\varepsilon _{s}\xi
_{A_{1}})(1+ik_{B_{2}}\xi _{A_{1}})^{-2}$. $J_{B_{2}A_{1}}^{\prime }$ is of
the same order of magnitude as $J_{A_{2}A_{1}}$. $J_{B_{2}A_{1}}^{\prime }$
does not create transitions from an extended state to a localized state. The
asymmetries in (\ref{ept}) and (\ref{tra2}) come from the different
separations (\ref{sl}) and (\ref{stadis}) of the single particle potential
energy for localized states and extended states. One should not confuse this
with the usual symmetry between transition probabilities for forward process
and backward process computed by the first order perturbation theory, where
two processes are coupled by the \textit{same} interaction. If we
approximate extended states $\xi _{B_{1}}$ and $\xi _{B_{2}}$ by plane waves
with wave vector $k_{1}$ and $k_{2}$, $K_{B_{2}B_{1}}\sim Z^{\ast
}e^{2}u\kappa ^{3}/(4\pi \epsilon _{0}\varepsilon _{s}[\kappa
+i(k_{2}-k_{1})])$, where $\kappa \sim (e^{2}/\epsilon _{0})(\partial
n/\partial \mu )$ is the Thomas-Fermi screening wave vector. In a lightly
doped or intrinsic semiconductor, $\kappa $ is hundreds or even thousands
times smaller than $1/a,$ $a$ is bond length. Since for most localized
state, localization length $\xi $ is several times $a$, therefore $%
K_{B_{2}B_{1}}\sim (\kappa \xi )^{2}K_{A_{2}B_{1}}^{\prime }$, is much
weaker than three other coupling constants.

\section{LE transition at low temperature}\label{lowLE}

The LE transition probability for low temperature $k_{B}T\leq \hbar
\overline{\omega }/10$ can be worked out as in (\ref{di}). Denote:%
\begin{equation}
f_{LE}(\omega _{\alpha })=\frac{1}{2}(\theta _{\alpha }^{A_{1}})^{2}csch%
\frac{\beta \hbar \omega _{\alpha }}{2},  \label{fle}
\end{equation}%
and the result is%
\[
W_{T}(A_{1}\rightarrow B_{2})=\frac{2\pi J_{B_{2}A_{1}}^{\prime 2}}{\hbar }%
\exp \{\frac{-\beta \Delta G_{LE}^{0}}{2}\}
\]%
\[
\exp \{-\frac{1}{2}\sum_{\alpha }(\theta _{\alpha }^{A_{1}})^{2}\coth \frac{%
^{\beta \hbar \omega _{\alpha }}}{2}\}\times
\]%
\[
\{\sum_{\alpha }f_{LE}(\omega _{\alpha })\frac{1}{2}[\delta (\Delta
G_{LE}^{0}+\hbar \omega _{\alpha })+\delta (\Delta G_{LE}^{0}-\hbar \omega
_{\alpha })]
\]

\[
+\sum_{\alpha\alpha^{\prime}}f_{LE}(\omega_{\alpha})f_{LE}(\omega
_{\alpha^{\prime}})\frac{1}{8}[\delta(\Delta G_{LE}^{0}+\hbar\omega_{\alpha
}+\hbar\omega_{\alpha^{\prime}})
\]

\[
+\delta(\Delta G_{LE}^{0}-\hbar\omega_{\alpha
}-\hbar\omega_{\alpha^{\prime}})
\]

\begin{equation}
+\delta (\Delta G_{LE}^{0}+\hbar \omega _{\alpha }-\hbar \omega _{\alpha
^{\prime }})+\delta (\Delta G_{LE}^{0}-\hbar \omega _{\alpha }+\hbar \omega
_{\alpha ^{\prime }})]\newline
+\cdots \}.  \label{nond}
\end{equation}

\section{EL transition at low temperature}\label{lowEL}

For low temperature ($k_{B}T\leq \hbar \overline{\omega }/10$), one can
expand the exponentials in (\ref{i1}) and (\ref{i2}) into power series. Then
the integrals can be carried out term by term. To 2-phonon processes, the
transition probability from extended state $|B_{1}\rangle $ to localized
state $|A_{2}\rangle $ is%
\[
W_{T}(B_{1}\rightarrow A_{2})=\frac{2\pi }{\hbar }\exp \{-\frac{\beta }{2}%
(E_{A_{2}}^{0}+\mathcal{E}_{A_{2}}^{b}-E_{B_{1}})\}
\]

\[
\exp[-\frac{1}{2}\sum_{\alpha}(\theta_{\alpha}^{A_{2}})^{2}\coth\frac{%
\beta\hbar\omega_{\alpha }}{2}]
\]

\[
\{\frac{1}{2}\sum_{\alpha^{\prime}}[\frac{(K_{A_{2}B_{1}}^{\prime%
\alpha})^{2}\hbar}{2M_{\alpha^{\prime}}\omega_{\alpha^{\prime}}}+\frac{%
(K_{A_{2}B_{1}}^{\prime\alpha}\Theta_{\alpha^{\prime}}^{A_{2}})^{2}}{4}\coth%
\frac {\beta\hbar\omega_{\alpha^{\prime}}}{2}]csch\frac{\beta\hbar\omega
_{\alpha^{\prime}}}{2}
\]

\[
\times\lbrack\delta(E_{A_{2}}^{0}+\mathcal{E}_{A_{2}}^{b}-E_{B_{1}}+\hbar%
\omega_{\alpha^{\prime}})+\delta(E_{A_{2}}^{0}+\mathcal{E}%
_{A_{2}}^{b}-E_{B_{1}}-\hbar\omega_{\alpha^{\prime}})]
\]

\[
+\sum_{\alpha^{\prime}}\frac{(K_{A_{2}B_{1}}^{\prime\alpha}\Theta
_{\alpha^{\prime}}^{A_{2}})^{2}}{8}(1-csch^{2}\frac{\beta\hbar\omega
_{\alpha^{\prime}}}{2})\sum_{\alpha^{\prime\prime}}f_{EL}(\omega
_{\alpha^{\prime\prime}})
\]

\[
[\delta(E_{A_{2}}^{0}+\mathcal{E}_{A_{2}}^{b}-E_{B_{1}}+\hbar\omega
_{\alpha^{\prime\prime}})+\delta(E_{A_{2}}^{0}+\mathcal{E}%
_{A_{2}}^{b}-E_{B_{1}}-\hbar\omega_{\alpha^{\prime\prime}})]
\]

\[
+\frac{1}{4}\sum_{\alpha^{\prime}\alpha^{\prime\prime}}[\frac{%
(K_{A_{2}B_{1}}^{\prime\alpha})^{2}\hbar}{2M_{\alpha^{\prime}}\omega_{%
\alpha^{\prime}}}+\frac{(K_{A_{2}B_{1}}^{\prime\alpha}\Theta_{\alpha^{%
\prime}}^{A_{2}})^{2}}{4}\coth\frac{\beta\hbar\omega_{\alpha^{\prime}}}{2}]
\]

\[
csch\frac{\beta \hbar\omega_{\alpha^{\prime}}}{2}f_{EL}(\omega_{\alpha^{%
\prime\prime}})
\]

\[
\times[\delta(E_{A_{2}}^{0}+\mathcal{E}_{A_{2}}^{b}-E_{B_{1}}+\hbar
\omega_{\alpha^{\prime}}+\hbar\omega_{\alpha^{\prime\prime}})+
\]

\[
+\delta(E_{A_{2}}^{0}+\mathcal{E}_{A_{2}}^{b}-E_{B_{1}}+\hbar\omega_{%
\alpha^{\prime}}-\hbar\omega_{\alpha^{\prime\prime}})
\]

\[
+ \delta(E_{A_{2}}^{0}+\mathcal{E}_{A_{2}}^{b}-E_{B_{1}}-\hbar\omega
_{\alpha^{\prime}}+\hbar\omega_{\alpha^{\prime\prime}})
\]

\[
+\delta(E_{A_{2}}^{0}+\mathcal{E}_{A_{2}}^{b}-E_{B_{1}}-\hbar\omega_{%
\alpha^{\prime}}-\hbar\omega_{\alpha^{\prime\prime}})]
\]

\[
+[\sum_{\alpha^{\prime}}\frac{(K_{A_{2}B_{1}}^{\prime\alpha}\Theta
_{\alpha^{\prime}}^{A_{2}})^{2}}{64}(2-csch^{2}\frac{\beta\hbar\omega
_{\alpha^{\prime}}}{2})]\times
\]

\[
\sum_{\alpha^{\prime\prime}\alpha^{\prime\prime\prime
}}f_{EL}(\omega_{\alpha^{\prime\prime}})f_{EL}(\omega_{\alpha^{\prime
\prime\prime}})
\]

\[
\times\lbrack\delta(E_{A_{2}}^{0}+\mathcal{E}_{A_{2}}^{b}-E_{B_{1}}+\hbar%
\omega_{\alpha^{\prime\prime}}+\hbar\omega_{\alpha^{\prime\prime\prime}})+
\]

\[
+\delta(E_{A_{2}}^{0}+\mathcal{E}_{A_{2}}^{b}-E_{B_{1}}+\hbar\omega
_{\alpha^{\prime\prime}}-\hbar\omega_{\alpha^{\prime\prime\prime}})
\]

\[
+\delta(E_{A_{2}}^{0}+\mathcal{E}_{A_{2}}^{b}-E_{B_{1}}-\hbar\omega
_{\alpha^{\prime\prime}}+\hbar\omega_{\alpha^{\prime\prime\prime}})+
\]

\[
\delta(E_{A_{2}}^{0}+\mathcal{E}_{A_{2}}^{b}-E_{B_{1}}-\hbar\omega
_{\alpha^{\prime\prime}}-\hbar\omega_{\alpha^{\prime\prime\prime}})]
\]

\[
-\sum_{\alpha^{\prime}}\frac{(K_{A_{2}B_{1}}^{\prime\alpha}\Theta
_{\alpha^{\prime}}^{A_{2}})^{2}}{16}csch^{2}\frac{\beta\hbar\omega
_{\alpha^{\prime}}}{2}
\]

\[
\lbrack \delta (E_{A_{2}}^{0}+\mathcal{E}_{A_{2}}^{b}-E_{B_{1}}+2\hbar
\omega _{\alpha ^{\prime }})\]
\begin{equation}
+\delta (E_{A_{2}}^{0}+\mathcal{E}%
_{A_{2}}^{b}-E_{B_{1}}-2\hbar \omega _{\alpha ^{\prime }})]+\cdots \},
\label{loweL}
\end{equation}%
where%
\begin{equation}
f_{EL}(\omega _{\alpha })=\frac{1}{2}(\theta _{\alpha }^{A_{2}})^{2}csch%
\frac{\beta \hbar \omega _{\alpha }}{2}.  \label{fel}
\end{equation}

%
%
%
%

\begin{thebibliography}{99}
\bibitem{eve} F. Evers and A. D. Mirlin, Rev. Mod. Phys. \textbf{80}, 1355
(2008).

\bibitem{dyr} J. C. Dyre and T. B. Schr\o der, Rev. Mod. Phys. \textbf{72},
873 (2000).

\bibitem{bern} B. Kramer and A. MacKinnon, Rep. Prog. Phys. \textbf{56},
1469 (1993).

\bibitem{ram} J. Rammer, Rev. Mod. Phys. \textbf{63}, 781 (1991).

\bibitem{pl} P. A. Lee and T. V. Ramakrishnan, Rev. Mod. Phys. \textbf{57},
287 (1985).

\bibitem{les} D. J. Thouless, Phys. Rep. \textbf{13}, 93 (1974).

\bibitem{mil} A. Miller and E. Abrahams, Phys. Rev. \textbf{120}, 745 (1960).

\bibitem{md} N. F. Mott and E. A. Davis, Electronic Processes in
Non-crystalline Materials, 2nd Ed., Clarendon Press, Oxford (1979).

\bibitem{kik} M. Kikuchi, J. Non-Cryst. Sol. \textbf{59}/\textbf{60}, 25
(1983).

\bibitem{gou} C. Gourdon and P. Lavallard, Phys. Stat. Sol. B\textbf{153},
641 (1989).

\bibitem{Over} H. Overhof and P. Thomas \textit{Electronic Transport in
Hydrogenated Amorphous Silicon} Springer Tracts in Modern Physics No. 114
(Springer, Berlin, 1989).

\bibitem{Yelon90} A. Yelon and B. Movaghar, Phys. Rev. Lett. 65, 618 (1990).

\bibitem{Yelon06} A. Yelon, B. Movaghar and R. S. Crandall, Reports on
Progress in Physics, 69, 1145 (2006).

\bibitem{wan} X. Wang, Y. bar-Yam, D. Adler and J. D. Joannopoulos, Physical
Review B 38, 1601 (1988).

\bibitem{cra} R. S. Crandall, Physical Review B 66, 195210 (2002).

\bibitem{Emin08} D. Emin, Physical Review Letters, 100, 166602 (2008).

\bibitem{sch} H. Scher and E. W. Montroll, Phys. Rev. B \textbf{12}, 2455
(1975).

\bibitem{ore} J. Orenstein and M. Kastner, Phys. Rev. Lett. \textbf{46},
1421 (1981).

\bibitem{kris} I. K. Kristensen and J. M. Hvam, Sol. Stat. Commun. \textbf{50%
}, 845 (1984).

\bibitem{mon} D. Monroe, Phys. Rev. Lett. \textbf{54}, 146 (1985).

\bibitem{marcu} R. A. Marcus, Rev. Mod. Phys., 65, 599 (1993).

\bibitem{Hol1} T. Holstein, Ann. of Phys., \textbf{8}, 325 (1959).

\bibitem{Hol2} T. Holstein, Ann. of Phys., \textbf{8}, 343 (1959).

\bibitem{rei} H. G. Reik, in J. T. Devereese (Ed), Polarons in Ionic and
Polar Semiconductors, Chapter VII, North-Holland/American Elsevier,
Amsterdam (1972).

\bibitem{Emin74} D. Emin, Phys. Rev. Lett. 32, 303 (1974).

\bibitem{Emin75} D. Emin, Adv. Phys. 24, 305 (1975).

\bibitem{Emin76} D. Emin and T. Holstein, Phys. Rev. Lett., \textbf{36}, 323
(1976).

\bibitem{Emin77} E. Gorham-Bergeron and D. Emin, Phys. Rev. B 15, 3667
(1977).

\bibitem{Emin91} D. Emin, Phys. Rev. B 43, 11720 (1991).

\bibitem{com} D. Emin: \textquotedblleft Aspects of Theory of Small Polaron
in Disirder Materials\textquotedblright, in \textit{Electronic and Sructural
Properties of Amorpous Semiconductors}, ed. p.261, by P. G. Le Comber and J.
Mort, Academic Press, London (1973).

\bibitem{Emin94} D. Emin and M.-N. Bussac, Phys. Rev. B 49, 14290 (1994).

\bibitem{bot} H. B\"{o}ttger and V. V. Bryksin, \textit{Hopping Conduction
in Solids}, VCH, Deerfield Beach, FL (1985).

\bibitem{and} P. W. Anderson, Rev. Mod. Phys. \textbf{50}, 191 (1978).

\bibitem{kubo} M.-L. Zhang and D. A. Drabold, Phys. Rev. B\textbf{81},
085210 (2010).

\bibitem{et} M.-L, Zhang, S.-S. Zhang and E. Pollak, Journal of Chemical
Physics, \textbf{119}, 11864 (2003).

\bibitem{kram} H. A.Kramers, Physica 7, 284 (1940).

\bibitem{dra} D. A. Drabold, P. A. Fedders, S. Klemm and O. F. Sankey, Phys.
Rev. Lett., \textbf{67}, 2179 (1991).

\bibitem{tafn} R. Atta-Fynn, P. Biswas and D. A. Drabold, Phys. Rev. B
\textbf{69}, 245204 (2004).

\bibitem{LLP} T. D. Lee, F. E. Low and D. Pines, Phys. Rev. \textbf{90}, 297
(1953).

\bibitem{big} Y. Pan, M. Zhang and D. A. Drabold, J. Non. Cryst. Sol.
\textbf{354}, 3480 (2008).

\bibitem{yue} Y. Pan, F. Inam, M. Zhang and D. A. Drabold, Phys. Rev. Lett.
100 206403 (2008).

\bibitem{sca} M.-L. Zhang, Y. Pan, F. Inam and D. A. Drabold, Phys. Rev. B
\textbf{78}, 195208 (2008).

\bibitem{tes} T. A. Abtew, M. Zhang and D. A. Drabold, Phys. Rev. B \textbf{%
76}, 045212 (2007).

\bibitem{cle} E. Clementi, D.L.Raimondi, and W.P. Reinhardt, J. Chem. Phys.
\textbf{47}, 1300 (1967).

\bibitem{si} K. W. B\"{o}er, Survey of Semiconductor Physics, Vol.1, John
Wiley, NewYork (2002).

\bibitem{force} H. Rucker and M. Methfessel, Phys. Rev. B \textbf{52}, 11059
(1995).

\bibitem{zal} R. Zallen, The Physics of Amorphous Solids, John Wiley and
Sons, New York. (1998).

\bibitem{ell} P. J. Elliott, A. D. Yoffe and E. A. Davis, in Tetradrally
Bonded Amorpous Semiconductors, p311, edited by M. H. Brodsky, S. Kirkpatric
and D. Weaire, AIP, NewYork (1974).

\bibitem{rog} R. E. DeWames, J. R. Waldrop, D. Murphy, M. Ray and R.
Balcerak, \textquotedblleft Dynamics of amorphous VO$_{x}$ for
microbolometer application\textquotedblright, preprint Feb. 5, 2008.

\bibitem{fren} J. Frenkel, Phys. Rev. \textbf{54}, 647-648, (1938).

\bibitem{abt} T. A. Abtew, M. Zhang, P. Yue and D. A. Drabold, J. Non-Cryst.
Sol. 354 2909 (2008).

\bibitem{eln} N. M. El-Nahass, H. M. Abd El-Khalek, H. M. Mallah and F. S.
Abu-Samaha, Eur. Phys. J. Appl. Phys. \textbf{45}, 10301 (2009).

\bibitem{sku} F. Skuban, S. R. Lukic, D. M. Petrovic, I. Savic and Yu. S.
Tver'yanovich, J. Optoelectronics and Advanced Materials \textbf{7}, 1793
(2005).

\bibitem{schm} H. Schmidt, M.Wiebe, B. Dittes and M. Grundmann, Appl. Phys.
Lett. \textbf{91}, 232110 (2007).

\bibitem{sag} P. Sagar, M. Kumar and R. M. Mehra, Sol. stat. Commun. \textbf{%
147}, 465 (2008).

\bibitem{winch} N.H. Winchell and A. H. Winchell, Elements of Optical
Mineralogy: Part 2, 2nd Ed. Wiley, New York (1927).

\bibitem{moni} A. H. Monish and E. Clark, Phys. Rev. B\textbf{11}, 2777
(1981).

\bibitem{elp} J. van Elp, H. Eskes, P. Kuiper and G. A. Sawatzky, Phys. Rev.
B\textbf{45}, 1612 (1992).

\bibitem{dew} R. Dewsberry, J. Phys. D: Appl. Phys. \textbf{8}, 1797 (1975).

\bibitem{motv} N. F. Mott, Phil. Mag., \textbf{19}, 835, (1969).

\bibitem{brod74} M. H. Brodsky and A. Lurio, Phys. Rev. \textbf{4}, 1646
(1974).

\bibitem{bah} S. K. Bahl and N. Bluzer, in Tetradrally Bonded Amorpous
Semiconductors, p320, edited by M. H. Brodsky, S. Kirkpatric and D. Weaire,
AIP, NewYork (1974).

\bibitem{sriv} J.K.Srivastava, M. Prasad and J. B. Wagner Jr., J.
Electrochem. Soc.: Solid-State Science and Technology \textbf{132}, 955
(1985).

\bibitem{awa} K. Awazu, J. Non-Cryst. Solids, \textbf{260}, 242 (1999).

\bibitem{oss} R. Ossikovski, B. Drevillon and M. Firon, J. Opt. Soc. Am. A%
\textbf{12}, 1797 (1995).

\bibitem{gmar} G. Marcano and R. Marquez, J. Phy. Chem. Sol. \textbf{64},
1725 (2003).

\bibitem{aiy} N. A. Jemali, H. A. Kassim, V. R. Deci, and K. N. Shrivastava,
J. Non-Cryst. Solids, \textbf{354}, 1744 (2008).

\bibitem{thou} D. J. Thouless, Phys. Rev. Lett. \textbf{39}, 1167(1977).

\bibitem{im} Y. Imry, Phys. Rev. Lett. \textbf{44}, 469 (1980).



\bibitem{oka} S. Okamoto, Y. Kanemitsu, Sol. Stat. Commun. \textbf{103}, 573
(1997).

\bibitem{kan} Y. Kanemitsu, M. Iiboshi and T. Kushida, J. Lumin. \textbf{87}-%
\textbf{89}, 463-465 (2000).


%
%
\end{thebibliography}
%

\end{document}